\documentclass[twocolumn]{aastex63}

\usepackage{multirow}
\usepackage{amsmath}
\usepackage{ulem}
\usepackage{lineno}

\newcommand{\Tcmb}{\mbox{$T_{\mbox{\tiny CMB}}$}}
\newcommand{\mass}{\mbox{$M_{\mbox{\scriptsize 500c}}$}}

\newcommand{\degs}{deg$^2$}

\newcommand{\muk}{\ensuremath{\mu {\rm K}}}

\newcommand{\spitzer}{{\sl Spitzer}}

\newcommand{\erosita}{{\sl eROSITA}}

\newcommand{\um}{$\mu$m}

\renewcommand{\vec}[1]{\mbox{\boldmath$#1$}}

\newcommand{\ncandfive}{266}  
\newcommand{\nconfirmfive}{244}  
\newcommand{\nconfirmfour}{448}  
\newcommand \nconfirmfourtofive{204}

\newcommand{\nstrong}{44}  
\newcommand{\areatot}{2770}
\newcommand{\areamasked}{122}  
  
\newcommand{\maskpct}{4.5}  

\newcommand{\nSN}{\ensuremath{\xi}}

\newcommand{\fsz}{\ensuremath{f_\mathrm{\mbox{\tiny{SZ}}}}}
\newcommand{\ysz}{\ensuremath{y_\mathrm{\mbox{\tiny{SZ}}}}}

\newcommand{\lcdm}{\ensuremath{\Lambda{\rm CDM}}}

\newcommand{\sptpol}{SPTpol}
\newcommand{\surveyname}{SPTpol Extended Cluster Survey}
\newcommand{\surveyshort}{SPT-ECS}

\newcommand{\nsummerpiscofourpfive}{173}

\newcommand{\nredmapper}{53,610}  
\newcommand{\nredmappervol}{21,092}  
\newcommand{\nredmapperspecs}{$\sim16,000$}  
\newcommand{\nredmapperspecsvol}{$\sim6,000$}  

\newcommand{\nredmapperspt}{652} 
\newcommand{\nredmapperzcut}{584} 
\newcommand{\nredmapperzcutvol}{410} 

\newcommand{\webaddress}{\url{http://pole.uchicago.edu/public/data/sptsz-clusters}}

\graphicspath{{./}{figures/}}

\shorttitle{SPTpol Extended Cluster Survey}
\shortauthors{Bleem et al.}  

\begin{document}

\title{The SPTpol Extended Cluster Survey}
\author[0000-0001-7665-5079]{L. E. Bleem} \affiliation{High Energy Physics Division, Argonne National Laboratory, 9700 South Cass Avenue, Lemont, IL 60439, USA} \affiliation{Kavli Institute for Cosmological Physics, University of Chicago, 5640 South Ellis Avenue, Chicago, IL 60637, USA}
\author[0000-0002-4900-805X]{S. Bocquet} \affiliation{Department of Physics, Ludwig-Maximilians-Universit\"{a}t,Scheinerstr.\ 1, 81679 M\"{u}nchen, Germany} \affiliation{High Energy Physics Division, Argonne National Laboratory, 9700 South Cass Avenue, Lemont, IL 60439, USA} \affiliation{Kavli Institute for Cosmological Physics, University of Chicago, 5640 South Ellis Avenue, Chicago, IL 60637, USA}
\author{B. Stalder} \affiliation{LSST, 933 North Cherry Avenue, Tucson, AZ 85721, USA} \affiliation{Center for Astrophysics $|$ Harvard \& Smithsonian, 60 Garden Street, Cambridge, MA 02138, USA}
\author{M. D. Gladders} \affiliation{Department of Astronomy and Astrophysics, University of Chicago, 5640 South Ellis Avenue, Chicago, IL 60637, USA } \affiliation{Kavli Institute for Cosmological Physics, University of Chicago, 5640 South Ellis Avenue, Chicago, IL 60637, USA}
\author{P. A. R. Ade} \affiliation{Cardiff University, Cardiff CF10 3XQ, United Kingdom}
\author{S. W. Allen} \affiliation{Kavli Institute for Particle Astrophysics \& Cosmology, P. O. Box 2450, Stanford University, Stanford, CA 94305, USA} \affiliation{Department of Physics, Stanford University, 382 Via Pueblo Mall, Stanford, CA 94305, USA} \affiliation{SLAC National Accelerator Laboratory, 2575 Sand Hill Road, Menlo Park, CA 94025, USA}
\author{A. J. Anderson} \affiliation{Fermi National Accelerator Laboratory, P. O. Box 500, Batavia, IL 60510, USA}
\author[0000-0002-0609-3987]{J. Annis} \affiliation{Fermi National Accelerator Laboratory, P. O. Box 500, Batavia, IL 60510, USA}
\author[0000-0002-3993-0745]{M. L. N. Ashby} \affiliation{Center for Astrophysics $|$ Harvard \& Smithsonian, 60 Garden Street, Cambridge, MA 02138, USA}
\author{J. E. Austermann} \affiliation{NIST Quantum Devices Group, 325 Broadway Mailcode 817.03, Boulder, CO 80305, USA}
% ...

\begin{abstract}

We describe the observations and resultant galaxy cluster catalog from the 2770 deg$^2$ \sptpol \ Extended Cluster Survey (\surveyshort).
Clusters are identified via the Sunyaev-Zel'dovich (SZ) effect, and confirmed with a combination of archival and targeted follow-up data, making particular use of data from the Dark Energy Survey (DES). 
With incomplete followup  we have confirmed as clusters \nconfirmfive\ of \ncandfive\ candidates at a detection significance $\xi \ge 5$  and an additional \nconfirmfourtofive \ systems at $4<\xi<5$. 
The confirmed sample has a median mass of $M_{500c} \sim {4.4 \times 10^{14} M_\odot h_{70}^{-1}}$ and a median redshift of  $z=0.49$, and we have identified  \nstrong \  strong gravitational lenses in the sample thus far.
Radio data are used to characterize contamination to the SZ signal; the median contamination for confirmed clusters is predicted to be $\sim$1\% of 
the SZ signal at the $\xi>4$ threshold, and $<4\%$ of clusters have a predicted contamination $>10\% $ of their measured SZ flux.
We associate SZ-selected clusters, from both \surveyshort\ and the SPT-SZ survey, with clusters from the DES redMaPPer sample, and find an offset distribution between the SZ center and central galaxy in general agreement with previous work, though with a larger fraction of clusters with significant offsets.  
Adopting a fixed \textit{Planck}-like cosmology, we measure the optical richness-to-SZ-mass ($\lambda-M$) relation and find it to be 28\% shallower than that from a weak-lensing analysis of the DES data---a difference significant at the 4 $\sigma$ level---with the relations intersecting at $\lambda=60$ .
The \surveyshort\ cluster sample will be particularly useful for studying the evolution of massive clusters and, in combination with DES lensing observations and the SPT-SZ cluster sample, will be an important component of future cosmological analyses.  \end{abstract}


\reportnum{DES-2019-0442}
\reportnum{FERMILAB-PUB-19-513-AE}

\section{Introduction} \label{sec:intro}
Clusters of galaxies, as tracers of the extreme peaks in the matter density field, are valuable tools for constraining cosmological and astrophysical models (see e.g.,  \citealt{voit04, allen11,weinberg13,kravtsov12} and references therein). 
Clusters imprint signals on the sky across the electromagnetic spectrum which have led to three main ways of observationally detecting these systems: as overdensities of galaxies in optical and/or near-infrared surveys 
(e.g., \citealt{abell58,koester07,wen12,rykoff14,bleem15a, eisenhardt08,wen18,oguri18,gonzalez19}), as sources of extended extragalactic emission at X-ray wavelengths (e.g., \citealt{gioia90,boehringer04,piffaretti11,ebeling10,mehrtens12,liu15b,adami18,klein19}),
and via their Sunyaev-Zel'dovich (SZ) signature \citep{sunyaev72} in millimeter (mm)-wave surveys. The latter two techniques rely on observables arising from the hot ($10^{7} -10^{8}$K) gas in the intracluster medium.   
While wide-field SZ-cluster selection is the newest realized technique---with the first cluster blindly detected in mm-wave survey data in 2008  \citep{staniszewski09}---the field has rapidly advanced with over 1,000 SZ-selected clusters published to date \citep{bleem15b,planck15-27,hilton18,huang19}.
SZ-selected cluster samples from high-resolution mm-wave surveys are of particular interest as they have low-scatter mass-observable proxies and, given the redshift-independence of the thermal SZ surface brightness, they are in principle mass-limited  \citep{carlstrom02,motl05}. 
Indeed, such samples have enabled SZ-cluster cosmological results that are competitive \citep{planck15-24,hasselfield13,bocquet19} with samples selected at other wavelengths (e.g., \citealt{vikhlinin09,mantz10a,mantz15}).

Cosmological constraints from samples of clusters are currently limited by an imperfect knowledge of both cluster selection and the connection of cluster observables to theoretical models.
The multi-wavelength nature of cluster signals allows for considerable opportunities to test and improve our understanding of these relations.  
Such explorations with SZ data and observations at other wavelengths can take many forms including:
(a) the use of optical, near-infrared, and X-ray data to both confirm SZ-cluster candidates and to provide empirical tests of models of SZ selection (e.g., \citealt{andersson11,planck12-1,planck12-4,liu15,bleem15b,planck15-27,hilton18,burenin18,barrena18}); (b) using SZ data to probe X-ray samples (e.g.,  \citealt{bender16b, czakon15, mantz16})
and to (c) test mass-optical observable scaling relations \citep{planck11-12,biesiadzinski12,sehgal13,rozo14,rozo15,mantz16,saro17,jimeno18}. 
Multi-wavelength observables are also used to constrain relevant quantities such as the spatial distribution of proxies for the cluster centers that feed into the derivation of such relations (e.g., \citealt{lin04b,george12,saro15, zhang19}).

In this work we expand the sample of SZ-selected clusters available for such studies using a new survey conducted using the \sptpol \ receiver  \citep{austermann12} on the South Pole Telescope (SPT). This wide and shallow survey complements the deeper surveys conducted with the SPT \citep{henning18,benson14} and will provide additional overlap for the comparison of cluster properties with the ACTPol \citep{debernardis16} and \textit{Planck} surveys. 
Here we present \ncandfive \  cluster candidates detected at a signal-to-noise $\xi>5$, \nconfirmfive \ of which are confirmed as clusters using optical and near-infrared data as well as via a search of the literature. We also report an additional \nconfirmfourtofive \  confirmed systems at $4<\xi<5$. 
Combining this dataset with the previously published SPT-SZ cluster sample  (\citealt{bleem15b}, hereafter B15), we use this expanded cluster sample to explore the SZ properties of massive optically selected clusters identified using the red-sequence Matched-Filter Probabilistic Percolation (redMaPPer) algorithm \citep{rykoff14} in the Dark Energy Survey Year 3 dataset.  

We organize this work as follows. 
In Section 2 we describe the survey observations and data reduction process. 
In Section 3 we describe the identification of cluster candidates including checks on the radio contamination of the sample and in Section 4 the cluster confirmation process including details on the external datasets used for this process.  
In Section 5 we present the full sample and several internal consistency checks with the SPT-SZ cluster sample. 
Detailed comparisons to the Dark Energy Survey redMaPPer sample including determination of the SZ-optical center offsets and SZ-mass-to-optical richness relation are presented in Section 6. 
We conclude in Section 7.

All optical magnitudes are quoted in the AB system \citep{oke74}. 
Except when noted, all masses are reported in terms of \mass, defined as the mass enclosed within a radius at which the average density is 500 times the critical density at the cluster redshift.
We assume a fiducial spatially flat $\Lambda$CDM cosmology with $\sigma_8=0.80$, $\Omega_b = 0.046$, $\Omega_m = 0.30$, $h = 0.70$, $n_s(0.002) = 0.972$, and $\Sigma m_\nu=0.06$ eV.
The normal distribution with mean $\vec{\mu}$ and variance $\vec\Sigma$ is written as $\mathcal{N}(\vec{\mu}, \vec\Sigma)$.  
Selected data reported in this work, as well as future updates to the properties of these clusters,  will be hosted at \webaddress. 

\section{Millimeter-wave Observations and Data Processing} \label{sec:Data}

\begin{figure*}[t]
\begin{center}
\includegraphics[width=7in]{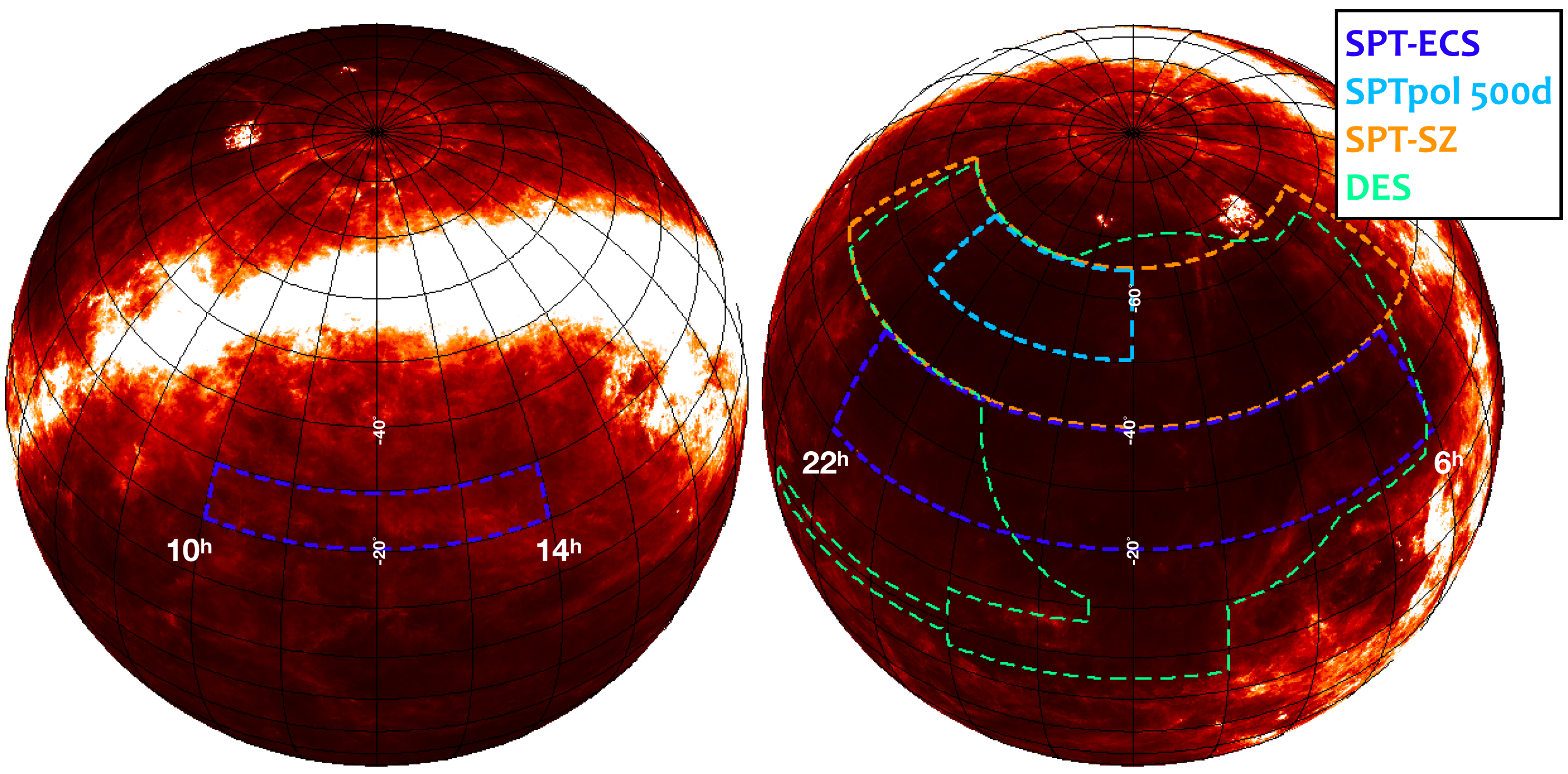}
\caption{Footprint of the \surveyname \ (dark blue) as compared to the SPT-SZ (orange) and \sptpol \ 500d survey (light blue). Optical-near infrared imaging from the Dark Energy Survey (green-dashed region) covers $\sim58\%$ of the survey footprint and is used to confirm a significant number of survey clusters presented in this work. The survey outlines are overlaid on the IRAS 100 $\mu$m  dust map  \citep{schlegel98} with the orthographic projection chosen such that the South Celestial Pole is at the top of the globe.  Beyond DES, \surveyshort \ also has significant overlap with the southern field of the Kilo-Degree Survey, the \textit{Herschel}--ATLAS survey, and the 2dFLenS spectroscopic survey.}

\label{fig:footprint}
\end{center}
\end{figure*}

The \surveyname \ (\surveyshort) is a 2770-square-degree survey that covers two separate regions of sky with low dust emission that lie north of previous areas surveyed using the SPT: a 2200 deg$^2$ region bounded in right ascension (R.A.) and declination ($\delta$) by  $22^h \leq$ R.A.$\leq 6^h$ and $-40^\circ <\delta <-20^\circ$ and a second 570 square degree region bounded by $10^h \leq$  R.A. $\leq  14^h$ and $-30^\circ \leq  \delta \leq -20^\circ$. 
These observations---conducted during the 2013, 2014, and 2015 austral summer months when data from the main 500-square degree \sptpol \ survey field (centered at R.A=$0^{h}$, $\delta=-57.5^\circ$, see \citealt{henning18}) would have been contaminated by scattered sunlight---serve to significantly increase the overlap of data from the SPT with that from other surveys including the Dark Energy Survey (DES; \citealt{flaugher15}), Kilo-Degree Survey (KIDS; \citealt{dejong13}), 2-degree Field Lensing Survey (2dFLenS; \citealt{blake16}), VISTA Kilo-Degree Infrared Galaxy Survey (VIKING, \citealt{edge13}), and \textit{Herschel}-ATLAS \citep{eales10}; see Figure \ref{fig:footprint}.  

\subsection{Observations}
The survey  was conducted using the \sptpol \ receiver that was installed on the 10 m South Pole Telescope \citep{carlstrom11} from 2012-2016. 
As detailed in \citet{austermann12}, the receiver is composed of 768 feedhorn-coupled polarization-sensitive pixels split between the two channels with 588  pixels at 150 GHz and 180 pixels at 95 GHz; 
each pixel contains two transition-edge-sensor bolometers resulting in 1536 detectors in total. 
The primary mirror is slightly under-illuminated resulting in beams well approximated by Gaussians with full width at half maximum of 1.2 and 1.7 arcmin at 150 and 95 GHz, respectively.

The survey is composed of ten separate $\sim250-270$ \degs \ ``fields", each imaged 
to noise levels of $\sim30-40$ \muk-arcmin at 150 GHz; see Table \ref{tab:fields}.
The fields were observed by scanning the telescope at fixed elevation back and forth in azimuth at $\sim0.55$ degrees/sec, stepping 10 arcmin in elevation, and then scanning in azimuth again. This process is repeated until the full field is covered in a complete ``observation". 
Each field was observed $>80$ times and twenty different dithered elevation starting points were used to provide uniform coverage in the final coadded maps. 

\subsection{Data Processing}\label{sec:processing}

The data processing and map-making procedures in this work follow closely those in previous SPT-SZ and \sptpol \ publications (see e.g., \citealt{schaffer11,bleem15b,crites15,henning18}). 
First, for each observation,  the time-ordered bolometer data (TOD) is corrected for electrical cross talk between detectors and  a small amount of bandwidth ($\sim1.4$~Hz and harmonics) is notch filtered to remove spurious signals from the pulse tube coolers that cool the optics and receiver cryostats. 
Next, using the cut criteria detailed in \citet{crites15}, detectors with poor noise performance, poor responsivity to optical sources, and/or anomalous jumps in TOD, are removed.
As this work is focused on temperature-based science we relax the requirement that both bolometers in a pixel polarization pair be active for an observation. 
Relative gains across the array are then normalized using a combination of regular observations of both an internal calibrator source and the galactic HII region RCW38. 
For the first field observed in the survey---\textsc{ra23hdec$-$35}\footnote{SPT fields are named for their central coordinates.}---the internal calibrator was inadvertently disabled during summer maintenance for $\sim50\%$ of the observations and so these data were relatively calibrated only with RCW38 observations. 

The TOD is then processed on a per-azimuth scan basis by fitting and subtracting a seventh-order Legendre polynomial, applying an isotropic common mode filter that removes the mean of all detectors in a 
given frequency, high-passing the data at  angular multipole  $\ell=300$ and low-passing the data at $\ell=20,000$. 
Sources detected in preliminary map making runs at $\ge 5\sigma$ ($\sim9-15$ mJy depending on field depth) at 150 GHz as well as bright radio sources detected in the Australia Telescope 20-GHz Survey (AT20G; \citealt{murphy10}) at the edges of the field are masked with a 4\arcmin \ 
radius during these filtering steps.  
The \surveyshort \ also contains a small number of sources with extended mm-wave emission (see Section \ref{sec:masking}) and more  conservative masks around these sources are applied in the filtering steps.\footnote{Given the arcminute scale beam, essentially all extragalactic sources at $z>0.05$ are unresolved in SPT data. See e.g., discussion of such sources in the SPT-SZ survey in W. Everett et al. (2019, in preparation). }
Following filtering, the TOD for each detector is then weighted based on the inverse noise-variance in the 1-3 Hz signal band and binned into $0$\farcm$25$ pixels in maps in the Sanson-Flamsteed projection \citep{calabretta02} using reconstructed telescope pointing. 
We have extended the characterization of the SPT pointing model to incorporate position information from all mm-wave-bright AT20G sources (typically 45-60 sources/field detected at S/N $>10$ were used compared to the 2-3 bright sources that proved sufficient in previous SPT analyses) to better constrain boom flexure and other mechanical aspects of the telescope at the elevations of these fields.  
With this extension we achieve reconstructed pointing performance of $\sim 3-4\arcsec$ root-mean-squared (rms) when comparing SPT source locations to AT20G positions. 

The single observation maps for each field are then characterized based on both noise properties and coverage; maps with significant outliers from the median of these distributions are flagged and excluded from the coaddition step.
The remaining maps are combined in a weighted sum based on their total pixel weights from the previous binning step; final maps consist of 78-150 observations per field. 

The \surveyshort \ fields were taken at significantly higher levels of atmospheric loading compared to other \sptpol \ survey data.\footnote{From 1.5--3 airmasses as compared to the median airmass of  $\sim1.2$  for the \sptpol \ main survey field.}
We found it necessary to augment our standard absolute calibration process (see e.g., \citealt{staniszewski09}) with two additional steps that make use of the 143 GHz full and half-mission temperature maps from the 2015 \textit{Planck} data release \citep{planck15-1,planck15-8}. 
The first step follows a similar method as the absolute temperature calibration conducted in previous SPT power spectrum analyses  (e.g, \citealt{hou18,henning18}).
We derive normalization factors to rescale each coadded map by first convolving the \textit{Planck} maps with the SPT beams and transfer functions (the latter resulting from the TOD filtering process described above) and the SPT maps with the \textit{Planck} beam and window function. 
Then, masking bright point sources in the field,  we set the normalization as the ratio from 900 $\le \ell \le 1600$ of the cross spectrum of the  \textit{Planck} half mission maps to the cross spectrum of the \textit{Planck} full mission map with the SPT maps.
The 95 GHz data required an additional calibration step as we found---especially in the fields centered at $\delta=-25^\circ$---that the responsivity of the detectors decreased with increasing airmass. 
This trend is well represented as a linear decline in sensitivity as a function of declination and we used the \textit{Planck} data to fit for and correct this variation across the fields.

\begin{deluxetable*}{ l c c c c c c}
\tablecaption{Summary information for the 10 fields that comprise the \areatot-square-degree \surveyname.\label{tab:fields}}
\tablewidth{0pt}
\tablehead{
\multicolumn{1}{l}{Name}\hspace{1.0cm} &
\multicolumn{1}{c}{\hspace{0.6cm}R.A.}\hspace{0.6cm}  &
\multicolumn{1}{c}{\hspace{0.6cm}$\delta$}\hspace{0.6cm} &
\multicolumn{1}{c}{\hspace{0.6cm}Area}\hspace{0.6cm} &
\multicolumn{1}{c}{$\sigma_{95}$} &
\multicolumn{1}{c}{$\sigma_{150}$} &
\multicolumn{1}{c}{\hspace{0.6cm}$\gamma_\textrm{field}$}\hspace{0.6cm}\\
\colhead{} &
\colhead{($^\circ$)} &
\colhead{($^\circ$)} &
\colhead{(\degs)} &
\colhead{(\muk-arcmin)} &
\colhead{(\muk-arcmin)}  &
\colhead{}  
}
\startdata
\textsc{ra23hdec-25} & 345.0 & -25.0 & 276.0 & 61.3 & 30.5 & 0.84 \\
\textsc{ra23hdec-35} & 345.0 & -35.0 & 250.2 & 59.4 & 36.6 & 0.80 \\
\textsc{ra1hdec-25} & 15.0 & -25.0 & 275.2 & 80.4 & 39.2 & 0.69 \\
\textsc{ra1hdec-35} & 15.0 & -35.0 & 251.8 & 61.5 & 36.6 & 0.79 \\
\textsc{ra3hdec-25} & 45.0 & -25.0 & 272.9 & 54.6 & 28.6 & 0.90 \\
\textsc{ra3hdec-35} & 45.0 & -35.0 & 248.8 & 43.8 & 25.3 & 1.04 \\
\textsc{ra5hdec-25} & 75.0 & -25.0 & 277.0 & 57.0 & 31.4 & 0.85 \\
\textsc{ra5hdec-35} & 75.0 & -35.0 & 250.3 & 54.8 & 31.6 & 0.88 \\
\textsc{ra11hdec-25} & 165.0 & -25.0 & 274.3 & 77.6 & 40.0 & 0.68 \\
\textsc{ra13hdec-25} & 195.0 & -25.0 & 270.8 & 50.7 & 30.0 & 0.90 \\
\enddata

\tablecomments{Listed are the field name, center,  source-masked effective area, and noise levels at both 95 and 150 GHz, as well as the ``field-renormalization'' factors discussed in Section \ref{subsec:szmass}. The survey contains an additional 122 square degrees that are masked in the cluster analysis owing to the presence of mm bright sources. Following \citet{schaffer11}, the noise levels are measured from $4000<\ell<5000$ using a Gaussian beam approximation with full-width at half maximum (FWHM) of 1.7 (1.2) arcmin at 95 (150) GHz respectively. The field renormalization factors are normalized with respect to the values from \citet{reichardt13} and \citet{dehaan16} for the SPT-SZ survey.}
\end{deluxetable*}

\section{Cluster Identification}\label{sec:extract}
Identification of cluster candidates in the \surveyshort \ proceeds in essentially identical fashion to previous SPT analyses (see, e.g., B15 for a recent example). 
This section provides an overview of the process; readers are referred to previous publications for more details.  

\subsection{Sky Model and Matched Filter}

The thermal SZ signal is produced by the inverse Compton scattering of CMB photons off high-energy electrons, such as those that reside in the intracluster medium of galaxy clusters.  
This produces a spectral distortion of the observed CMB temperature at the location \textbf{x} of clusters given by the line-of-sight integral \citep{sunyaev72}: 

\begin{equation}
\begin{split}
\Delta T(\mathbf{x},\nu) &= \Tcmb \ \fsz(\nu)\int n_\mathrm{e}(\mathbf{r}) \frac{k_\mathrm{B}T_\mathrm{e}(\mathbf{r}) }{m_\mathrm{e}c^{2}} \sigma_\mathrm{T} dl  \\
  &\equiv  \Tcmb \ \fsz(\nu) \ \ysz(\mathbf{x})
\end{split}
\end{equation}
where  $\Tcmb=2.7260\pm 0.0013$~K is the mean CMB temperature \citep{fixsen09b}, 
$\fsz(\nu)$ is the frequency ($\nu$) dependence of the thermal SZ effect \citep{sunyaev80}, 
$n_\mathrm{e}$ the electron density, $T_\mathrm{e}$ the electron temperature,
$k_\mathrm{B}$ the Boltzmann constant,
 $m_\mathrm{e} c^2$ the electron rest mass energy, $\sigma_\mathrm{T}$ the Thomson cross-section, 
 and $\ysz$ is the Compton $y$-parameter.
This effect results in a decrement at the two channels measured by the \sptpol \ receiver; 
for a non-relativistic thermal SZ spectrum the effective band centers are 95.9 and 148.5~GHz.\footnote{
Though see e.g., \citet{wright79,nozawa00,itoh04,chluba12} for discussion of relativistic corrections which become relevant at \mbox{$T_\mathrm{e} \gtrsim 8 \ \mathrm{keV}$}.}

To identify candidate galaxy clusters we use a spatial-spectral filter designed to optimally extract thermal SZ cluster signals \citep{melin06}. 
This ``matched-filter" approach has been widely used in both previous SPT publications as well as in analyses by other  experiments (see e.g., \citealt{planck15-27,hilton18}). 
We model the cluster profile as a projected spherical $\beta$-model with $\beta$ fixed to 1 \citep{cavaliere76}: 
\begin{equation}
\Delta T = \Delta T_{0}(1+ \theta^{2}/\theta_\mathrm{c}^{2})^{-(3\beta-1)/2}
\label{eqn:beta}
\end{equation}
where the normalization $\Delta T_{0}$ is a free parameter and the core radius, $\theta_\mathrm{c}$, is allowed to vary in twelve equally spaced steps from  $0\farcm25$ to 3\arcmin.

\subsection{Masking}
\label{sec:masking}
To prevent spurious decrements from the filtering process we mask regions around bright emissive sources before applying the matched filters to the maps.
These sources are detected in the 150~GHz data using a matched filter designed to optimize the signal-to-noise of point sources.
Masks of 4\arcmin \ radius are placed over sources detected at $>5\sigma$  
and candidates detected within 8\arcmin \ of these sources are excluded from the final cluster lists.  
Additionally, as referenced above in Section \ref{sec:processing}, there are three extended sources in these fields---NGC 55, 253, and 7293 \citep{dreyer88}  
and  one exceptionally bright quasar---QSO B0521-365 (e.g., \citealt{planck18-54})---that require additional masking.
Masks of radius $0.33^\circ$ are used for the NGC sources and radius $0.25^\circ$ for the quasar.
Regions around these sources are also inspected following the cluster filtering process and a small number of spurious candidates are rejected. 
In total \areamasked \ \degs \ are masked, \maskpct\%  of the full survey area.

\subsection{Candidate Identification} \label{subsec:cand}
Cluster candidates are identified as peaks in the matched-filtered maps.  
For each location we define our SZ observable, $\xi$, as the maximum detection-significance over the twelve filter scales.
As in prior SPT analyses, there is a small declination dependence in the noise owing to atmospheric loading, detector responsivity, and coverage changes across each field. 
To capture this in the $\xi$ estimates, each filtered map is split into 90\arcmin \ strips in declination and---as in \citet{huang19}---noise in each strip is measured by measuring the standard deviation of a Gaussian fit to unmasked pixels. 
In this work, all candidates $\xi \ge 5 $ are reported, and for $4<\xi<5$, where our followup is currently highly spatially incomplete, we also report confirmed systems  in the DES common region (see Section \ref{sec:confirm}).

\subsection{Field Depth Scaling and False Detection Rate}\label{subsec:false}
We make use of simulations to estimate the contamination of our catalogs by spurious detections and to renormalize the  
measured SZ detection significances to account for the varying field depths (see Section \ref{subsec:szmass}).
Simulations were previously used to this effect in e.g., \citet{reichardt13}, \citet{dehaan16}, \citet{huang19}; we briefly overview the process here and describe some small changes to the process from the SPT-SZ simulations.  
For more details on these simulations see \citet{huang19}. 

For each field we construct sets of simulated mm-wave skies consisting of:
\begin{itemize}
    \item primary lensed CMB \citep{keisler11}.
    \item signals from Poisson and clustered dusty sources that we approximate as Gaussian random fields with amplitude and spectral indices matching \citet{george15}.
    \item discrete radio sources below the masking threshold with the source population drawn from the model of \citet{dezotti05} with spectral indices drawn from the results of \citet{george15} and \citet{mocanu13}.
    \item thermal SZ constructed using a halo light cone from the Outer Rim \citep{habib16, heitmann19} simulation with thermal SZ profiles painted for each halo with $M_{200c}>10^{13}$ following the methodology of \citet{flender16} and using the pressure profiles of \citet{battaglia12}.  The thermal SZ power is consistent with the results of \citet{george15}. The SZ signal is omitted in the false detection simulations.  
    \item atmospheric and instrumental noise from jackknife noise maps constructed via coadding field observations where half of the observations were randomly multiplied by $-1$. 
\end{itemize}

Each sky realization is convolved with the SPT beam and transfer function.
As in \citet{huang19}, there are two significant changes compared to simulations used for SPT-SZ cluster studies.  First, we use discrete radio sources, as opposed to Gaussian random fields, to account for radio contamination. This change was found to be important for properly capturing the false detection rates of the deeper SPTpol 100d and 500d cluster surveys but has negligible impact at the noise levels of the \surveyshort \ and SPT-SZ surveys. We adopt it for consistency here. Second, we use the measured SPT beams, as opposed to Gaussian approximations, which enables more consistent scalings between the SPT-SZ and SPTpol experiments.

To estimate the number of spurious detections in each field, we run the cluster detection algorithm on the simulated SZ-free maps. As in \citet{dehaan16}, to reduce shot noise in our estimates from our finite number of simulations, we model the false detection rate with the function: 
\begin{equation}
    N_\textrm{false}(>\xi) = \alpha_\textrm{field}e^{-\beta_\textrm{field} \left( \xi-5 \right)} \times \text{field area}
\end{equation}
All of the fields are well approximated by $\alpha\sim0.008$ and  $\beta=4.3$; as each field is approximately 260 square degrees this results in $\sim 2$ false detections/field expected above $\xi=5$ and $17-18$ above $\xi=4.5$.

As detailed in \citet{dehaan16}, the  field depth rescaling factors, which track changes in ``unbiased significance'' as a function of mass for the varying field depths, are determined by measuring the signal-to-noise of simulated clusters at their known locations and optimal filter scales from the simulated maps (see also \ref{subsec:szmass}). 
We list the field depth rescaling factors $\gamma_\mathrm{field}$ in Table~\ref{tab:fields}. Following previous SPT publications, the absolute normalization is set to correspond to the unit scaling adopted in \cite{reichardt13}.
While in principle the field scaling simulations should be sufficient to properly scale the \surveyshort \ field depths relative to SPT-SZ, the extra calibration steps required for the \surveyshort \ survey make this challenging. To capture any residual uncertainty in this process we introduce a new parameter, $\gamma_\mathrm{ECS}$, which rescales all field scalings in the \surveyshort \ survey $\gamma_{\mathrm{SPT-ECS},i}=\gamma_\mathrm{ECS}\times\gamma_{\mathrm{field},i}$. With this parametrization, $\gamma_\mathrm{ECS}=1$ means that our simulations capture the entirety of the relative difference in effective depth between SPT-SZ and \surveyshort. We empirically calibrate $\gamma_\mathrm{ECS}$ in sections~\ref{sec:abundance} and \ref{subsec:lambdam}.

\subsection{Potential Contamination of the SZ Sample From Cluster Member Emission}
\label{sec:contam}
Galaxy clusters contain an overdensity of galaxies relative to the field, and galaxies emit
radiation at mm wavelengths. Since the thermal SZ signal from the cluster gas is a
decrement in the frequency bands used in this work, any positive emission above the background
will act as a negative bias to the SZ signature. We can classify the potential bias from cluster galaxy
emission into two regimes, one in which the integrated emission from many cluster members produces
an average bias to all clusters in a given mass and redshift range,
with little variation from cluster to cluster; and one in which a single bright galaxy
(or a very small number of bright galaxies) imparts a significant bias to a random subsample of clusters.
We can also separate the contributions to this effect from the two primary classes of mm-wave-emissive
sources: active galactic nuclei producing synchrotron emission (``radio sources'') and star-forming
galaxies producing thermal dust emission (``dusty sources'').

The contribution to the second type of bias from dusty sources is expected to be negligible, 
because the dusty source population falls off steeply at high flux \citep[e.g.,][]{mocanu13}, so that
the areal density of dusty sources bright enough to fill in a cluster decrement at a level important for this work 
is very low. This statement is for the field galaxy population, so if galaxies in clusters were more likely than field galaxies to be dusty and star-forming, 
the bright population could still be an issue. 
In fact, the opposite is expected to be true; i.e., compared to the field population, galaxies in clusters are less likely to be dusty and star-forming, at least at $z <$ 1 (e.g., \citealt{bai07,brodwin13,alberts16}).
In \citet{vanderlinde10}, it was argued 
that the other regime of bias from dusty sources
is also negligible for clusters more massive than $\sim$2$\times 10^{14} M_\odot$, 
which includes all the clusters in this sample (see also, e.g., \citealt{soergel17} for an analysis of a sample of low-$z$ optically selected clusters, and \citealt{erler18},  \citealt{melin18} for explorations of the \textit{Planck} sample).  

To assess the potential contamination from radio sources, we make use of the publicly 
available maps from the 1.4~GHz 
National Radio Astronomy Observatory (NRAO) Very Large Array (VLA) Sky Survey
(NVSS, \citealt{condon98}).\footnote{Maps downloaded via anonymous ftp from  
https://www.cv.nrao.edu/nvss/postage.shtml}
NVSS covers the full sky north of declination $-40^\circ$ and thus has nearly 100\% 
overlap with the survey fields in this work. The data for the NVSS were taken between
1993 and 1997, so source variability will limit the fidelity of the estimate of
contamination to any individual cluster, but we can make some statements about the
average or median contamination across the catalog and the fraction of clusters 
expected to be strongly affected by radio source contamination.

For each of our survey fields, we download 
all NVSS postage-stamp maps (each 4 $\times$ 4 degrees) that have any overlap with that 
field and reproject them onto the same pixel grid as used in our cluster analysis. We then
make beam- and transfer-function-matched NVSS maps for each of the SPTpol observing
frequencies by convolving the NVSS maps with a kernel defined by the Fourier-space ratio
of the SPTpol beam and transfer function at that frequency and the effective NVSS beam
(a 45-arcsec FWHM Gaussian). We scale the intensity in these maps from 1.4~GHz to SPTpol frequencies
assuming a spectral index of $-0.7$ (roughly the mean value found for radio sources in clusters
by \citealt{coble03}), and we convert the result to CMB fluctuation temperature.

We then combine the SPTpol-matched NVSS maps in our two bands using the 
same band weights as used in the cluster-finding process (Section~\ref{sec:extract}) and 
filter the output with the same $\beta$-model-matched filters as used in the cluster-finding
process. For each cluster candidate in the catalog, we take the combined NVSS 
map filtered with the same $\beta$-model profile as the cluster candidate, and we 
record the value of the combined, filtered NVSS map at the candidate location. We divide
that value by the same noise value used in the denominator of the $\xi$ value for the cluster
candidate, and we record that value as our best estimate of the contamination to the $\xi$
value of that cluster candidate from radio sources. Since the NVSS maps contain all the
radio flux at 1.4~GHz (not just the sources bright enough to be included in the NVSS catalog),
this test accounts for both regimes of bias discussed above.

The median contamination calculated in this way is $\Delta \xi_\mathrm{med} = -0.05$, or
1\% of the threshold value for inclusion in the catalog of $\xi=5$.
Of the 266 candidates in the catalog, 
13 ($\sim$5\%) have a predicted contamination of greater than 10\% of their measured SZ flux, and 
7 ($\sim$2.6\%) have a predicted contamination of greater than 20\%.
One cluster candidate, 
SPT-CL~J2357-3446, has an anomalously large predicted bias of $\Delta \xi = -11.1$.
This candidate is almost certainly the low-redshift ($z=0.048$) cluster Abell 3068 
(separation $0\farcm1$), and it is within $0\farcm6$ of the NVSS source 
NVSS J235700-344531, which has a catalog 1.4~GHz flux of 1.28 Jy. This NVSS
source is a cross-identifcation of PKS 2354-35, which lies at a redshift consistent
with being a member of A3068 ($z=0.049$) and is identified as the central galaxy of
this cluster by many authors \citep[e.g.,][]{schwartz91}. Given the relative redshift 
dependence of the thermal SZ signature of clusters and the flux density of member 
emission, it is not surprising that the highest level of radio source contamination occurs
in one of the lowest-redshift clusters in the sample. It is somewhat surprising, though, 
that a cluster with a predicted radio source contamination of $\Delta \xi > 10$ would 
be detected at $\xi = 5.5$, as this one is in our catalog. The apparent answer to this
puzzle is source variability. More recent observations of this source with the 
Australia Compact Telescope Array (ATCA) at 5~GHz \citep{burgess06} resulted in a
measured flux density of 99 mJy, which would imply a spectral index of $<-2.0$ if 
naively combined with the NVSS measurement at 1.4~GHz. We conclude that during 
our observations, the 150~GHz flux of this source was likely $<10$ mJy
(as implied by the ATCA measurement and a spectral index of $-0.7$) rather than the 
$\sim$50 mJy implied by the NVSS measurement.
Finally, we also note that all previous SPT cluster cosmology
results have cut clusters below $z=0.3$ or $0.25$, so this cluster would not normally 
be included in an SPT cosmology analysis.

These contamination numbers will be diluted somewhat by any false detections.
However, removing the 22 unconfirmed candidates at $\xi>5$ has negligible effect. 
If we extend the sample to all confirmed systems at $\xi>4$ (for a total of \nconfirmfour \ clusters), we find a similar median contamination ($\Delta \xi_\mathrm{med} = -0.050$)
and fraction of systems above a given level of 
contamination: 17 ($\sim$4\%) 
and 8 ($\sim$2\%) above 10\% and 20\% contamination, respectively.
We flag candidates with $>10\%$ potential contamination of their measured SZ signal in Tables \ref{tab:sampletable} and \ref{tab:sample4sigma}. 

\subsection{Clusters in Masked Regions}\label{subsec:maskedclusters}
\label{sec:maskedregion}
In addition to the potential bias to our sample from the mm-wave emission from cluster members, 
there is a potential bias from the avoidance of mm-wave-bright sources in our cluster-finding.
As discussed in Section~\ref{sec:masking}, we discard any cluster detection within 8$^\prime$ of a
source detected at 5$\sigma$ ($\sim$9-15~mJy, field-dependent) at 150~GHz. 
If there were a strong physical association between galaxy clusters and such sources, our measured cluster 
abundance would be biased low.

The majority of sources with 150~GHz flux density above 9 mJy are flat-spectrum quasars (see e.g., \citealt{mocanu13,gralla19}), and, based on studies of radio galaxies from lower frequency surveys (e.g., \citealt{lin09,gralla11,gralla14,gupta17}), there is not expected to be a significant SZ selection bias from these sources.  
However, we can perform several checks on the effects of masking with the data in hand. 
First, as in B15, we perform a secondary cluster search, this time only masking emissive sources detected at $>~100$ mJy at 150 GHz.  
Each candidate from this run was visually inspected and, as expected, this candidate list was dominated by filtering artifacts; no new clusters were identified in this secondary run.

We also check for any statistical association between the flux-limited DES redMaPPer optically selected cluster catalog and associated random locations (discussed in more detail below in Section \ref{sec:des} and in \citealt{rykoff16}) and SPT-selected emissive sources.  To increase the sensitivity of this test, we also include sources from the SPT-SZ survey, which had a 5$\sigma$ source threshold of lower flux ($\sim$6 mJy; W. Everett et al., (2019, in preparation)).
We first measure the probability of optical clusters and random locations to be within the 8\arcmin \ source masks and find marginal differences between the two. 
Adopting a 3\arcmin \ radius to reduce the noise from chance associations and further restricting the cluster sample to $z>0.25$ where we expect the SPT selection to be well understood, we find an excess probability over random of $\lesssim 1\%$ for clusters to fall in the source-masked regions (see Table \ref{tab:pscut}). While the purity of the flux-limited sample is expected to decrease at high redshift (thus limiting our ability to test for trends with redshift), we note that we find no significant difference in the fraction of clusters in masked regions between the full sample and two subsamples constructed by splitting the optical sample at its median redshift  of $z=0.755$.

\begin{deluxetable}{lcccccc}[t]  
\tablecolumns{5}
\tablecaption{\\ Optical Clusters near mm-wave bright sources\label{tab:pscut}}
\tablehead{ 
\colhead{$\lambda$ range} &
\colhead{$N_\textrm{clusters}$} & 
\colhead{\% in masked region} &
\colhead{$N_\textrm{clusters}^{\textrm{SPT-SZ}}$} & 
\colhead{\%}}
\startdata
randoms & 2.1e6 &  0.9 & 1.2e6 & 1.05\\
20$-$30 & 2.3e4 &1.11 & 1.3e4 &  1.4\\
30$-$50 & 9.3e3 & 0.95 & 5.4e3 & 1.1\\
50$-$80 & 1.9e3 &  1.22 & 1.1e3 &  1.65\\
$>80$ & 3.5e2 & 1.13 &  2.0e2 & 1.5\\


\enddata 
\tablecomments{Percentage of DES redMaPPer clusters at $z>0.25$ that fall within 3\arcmin \ of bright emissive sources; $N_\textrm{clusters}$ corresponds to the total number of clusters in a given richness bin within the SPT-SZ+\surveyshort \ (Left) or SPT-SZ only (Right) footprint. The top row provides statistics for random sources. Less than 1\% of clusters over random fall in the masked areas.} 
\end{deluxetable}

\section{External Datasets and Cluster Confirmation}\label{sec:confirm}

To confirm the SZ candidates as galaxy clusters we make use of  targeted optical and near-infrared follow-up observations,  data drawn from the wide area 
Dark Energy Survey \citep{flaugher15}, the Pan-STARRS1 survey \citep{chambers16}, the all sky Wide-field Infrared Survey Explorer (WISE) dataset \citep{wright10}, and the literature. 
In this section we describe each dataset and how it is used to confirm and/or characterize the SZ-selected clusters. 
Overall, we focus our targeted follow-up efforts on ensuring nearly complete imaging of high-significance ($\xi>5$) cluster candidates to depths sufficient to confirm clusters to $z\sim 0.8-1.0$. 
For lower-significance targets we rely significantly more on the availability of wide-area imaging datasets. 

\subsection{External Datasets}
 
\subsubsection{The Dark Energy Survey and redMaPPer}\label{sec:des}
The Dark Energy Survey  is a recently completed ${\sim5000}$ \degs \ optical-to-near-infrared imaging survey conducted with the DECam imager \citep{flaugher15} 
 on the 4m Blanco Telescope at Cerro Tololo Inter-American Observatory.
The survey was designed to have significant overlap with the original SPT-SZ survey (see Figure \ref{fig:footprint}) and we have increased this overlap with the addition of  \surveyshort. 
In this work we make use of the DES data acquired in years 1-3 of the survey; this data reaches signal-to-noise 10 in 1\farcs95 apertures in the \textit{grizY} bands at [24.33, 24.08, 23.44, 22.69, 21.44] magnitudes with resolution---given by the median FWHM of the point spread functions---of [1.12, 0.96, 0.88, 0.84, 0.9] arcseconds respectively \citep{abbott18b}\footnote{Data available: \url{ https://des.ncsa.illinois.edu/releases/dr1}}. 

We make particular use of the  redMaPPer optically selected galaxy cluster sample drawn from the DES data. 
As its name implies, redMaPPer (hereafter RM) is a red-sequence-based cluster finder that identifies clusters as overdensities of red galaxies based upon galaxy positions, colors, and brightness \citep{rykoff14,rykoff16}. 
Each RM cluster detection provides---amongst other quantities---a cluster redshift, a probabilistic center (based on the consistency of bright cluster members with the observed properties of cluster central galaxies), a similarly probabilistic cluster member catalog, and a total optical richness,  $\lambda$, that is the sum of all the cluster member probabilities corrected for various masking and completeness effects. 
The RM sample has been shown to have excellent redshift precision, with uncertainties in redshift estimates of $\sigma_z/(1+z)\sim 0.01$ for clusters $z<0.9$. 

There are 2 different RM catalogs: a ``flux-limited'' sample that includes significant numbers of high-redshift clusters for which the optical richness estimates must be extrapolated and  a ``volume-limited" sample for which the DES data is sufficiently uniform and deep that the richnesses can be well measured; DES cluster cosmology constraints are derived using the volume-limited sample \citep{rykoff16,mcclintock19}. 
In this work, we explore characteristics of the joint SPT-RM cluster sample using the volume-limited catalog.\footnote{RM catalog version 6.4.22}
In total there are \nredmapper \ (\nredmappervol ) RM clusters at $\lambda >20$ in the full (volume-limited) DES sample, with $\sim36,000$ (\nredmapperspecs) and $\sim14,000$ (\nredmapperspecsvol) of these clusters within the total SPT and \surveyshort \ survey area, respectively.

\subsubsection{The Parallel Imager for Southern Cosmology Observations}

We use the Parallel Imager for Southern Cosmology Observations  (PISCO; \citealt{stalder14})---a new imager with a 9\arcmin \ field-of-view installed on the 6.5 m Magellan/Clay telescope at Las Campanas Observatory in Chile---to obtain approximately uniform depth \textit{griz'} imaging data  for over 500 SPT-selected clusters and cluster candidates, including \nsummerpiscofourpfive \ candidates at $\xi\geq 4.5$ in \surveyshort. 
These data were obtained as part of an ongoing effort to characterize the strong lensing and bright galaxy populations of the SPT cluster sample.  

To analyze the PISCO data, we have constructed a reduction pipeline that includes standard corrections (i.e., overscan, debiasing, flat-fielding, illumination, though we note defringing is not required) as well as specialized routines that correct the data for non-linearities and artifacts caused by bright stars. 
After the images have been flatfielded, we use the PHOTPIPE pipeline  \citep{rest05a, garg07,miknaitis07} for both astrometric and relative calibration prior to coaddition.
We make use of stars from the DES DR1 public release  \citep{des18}, from the second \textit{Gaia} data release \citep{gaia18}, or from the Pan-STARRS 1 release  \citep{flewelling16}  to obtain sufficient numbers of sources for  good astrometric solutions.  
Images are coadded using  SWarp \citep{bertin02} and sources are detected in the coadded images using SExtractor \citep{bertin96}  in dual-image mode with the $r-$band image set as the detection image. 
We find the typical $\sim85\%$ completeness depth of these images to be $r=24.3$.  
We separate bright stars and galaxies using the  \textit{SG} statistic  \citep{bleem15a} and use these stars to calibrate the photometry with  stellar locus regression \citep{high09};  absolute magnitudes are set using the 2MASS point source catalog \citep{skrutskie06}.

\subsubsection{\spitzer /IRAC and Magellan/Fourstar}
Based on initial PISCO imaging, we were able to identify a small number of high-redshift cluster candidates worthy of additional follow-up observations. 
Two systems 
were imaged as part of a \spitzer \ Cycle 11 program and 5 additional \surveyshort \ cluster candidates were part of a Cycle 14 program (ID: 11096,14096; PI:Bleem)\footnote{Data available: \url{https://sha.ipac.caltech.edu/applications/Spitzer/SHA/}}. 
The Cycle 11 (14) candidates were observed with \spitzer /IRAC \citep{fazio04} for 360 (180) s on source time integration time in both the 3.6 and 4.5 \um \ bands. 
These data were reduced as in \citet{ashby09}, and are of sufficient depth for cluster confirmation to $z\sim 1.5$. 

We have additionally obtained ground-based near-infrared $J-$band imaging for 19 candidates with the Fourstar imager \citep{persson13} installed on the Magellan/Baade telescope. 
For each candidate, a large number of short exposures were taken using predefined dither macros provided in the instrument control software with the candidate centered on one of the four Fourstar detectors. 
These images were flatfielded using IRAF routines \citep{tody93}, astrometrically registered and relatively calibrated, and coadded using the PHOTPIPE pipeline. 
Coaddition was performed with {\tt SWarp} and source identification with {\tt SExtractor}.
Absolute calibration is tied to 
the J-band flux from stars in the 2MASS point source catalog. 
While conditions varied between the Fourstar observations these data are typically sufficient to confirm clusters to $z\sim1.2$ or better.

\subsubsection{Pan-STARRS1}

The SPT fields north of $\delta= -30^\circ$ have been imaged by the Panoramic Survey Telescope and Rapid Response System  (Pan-STARRS) in the \textit{grizy$_p$} filter bands as part of the Pan-STARRS 3$\pi$ Steradian Survey \citep{chambers16}. 
In this work we make use of the first data release \citep{flewelling16} available for download from the Mikulski Archive for Space Telescopes.\footnote{\url{http://panstarrs.stsci.edu/}}
This data release contains images and source catalogs from the ``stack'' coadded image products. 
These data are shallower than both the DES survey data  and our targeted follow-up imaging, with 5$\sigma$ point source depths of [23.3, 23.2, 23.1, 22.3, 21.4] magnitudes in the  \textit{grizy$_p$} bands respectively. 
Exploring the properties of clusters in the Pan-STARRS footprint that we confirmed in DES, PISCO, and the literature (see below), we find the Pan-STARRS data typically enables robust confirmation of clusters to $z<0.6 -0.7$. 

\subsubsection{WISE}
As demonstrated in e.g., \citet{gonzalez19}, observations from the Wide-field Infrared Survey Explorer (WISE; \citealt{wright10}) are an excellent resource for identifying high-redshift clusters. 
Of particular relevance for this work are the observations in the [W1] and [W2] filter bands at 3.4$\mu$m and 4.6$\mu$m which we use to confirm cluster candidates by identifying overdensities of high-redshift galaxies at a common 1.6 $\mu$m rest frame (see e.g., \citealt{muzzin13b}). 
Here we make use of ``unWISE'' a new processing of WISE and NEOWISE \citep{mainzer14} data that reaches 3 times the depth of the AllWISE data \citep{meisner17,schlafly19}\footnote{Data available: \url{http://unwise.me/imgsearch/}}.

\subsubsection{Literature Search}\label{subsec:lit}
We additionally search the literature for known clusters in the vicinity of the SZ-selected candidates. 
Using the SIMBAD\footnote{http://simbad.u-strasbg.fr/simbad} database we search for  systems within a  5\arcmin \ radius of the candidate locations. 
When such a system is found we consider it to be a match if it is at $z<0.3$; we reduce the matching radius to 2\arcmin \ for clusters at higher redshifts (except for systems in the \textit{Planck} catalog, see Section \ref{subsec:planckc} below).
When available, we adopt reported spectroscopic redshifts as the SPT cluster redshifts for previously identified systems. 
We also make use of reported photometric redshifts for a small number of systems not in DES, Pan-STARRS1, or directly targeted in our follow-up imaging. 
When possible we use the other external datasets to identify spurious associations with previously reported systems, finding several in this distance-based match. 

\subsection{Cluster Confirmation}
\begin{figure}[t]
\hspace*{-0.25in}
\includegraphics[width=3.5in]{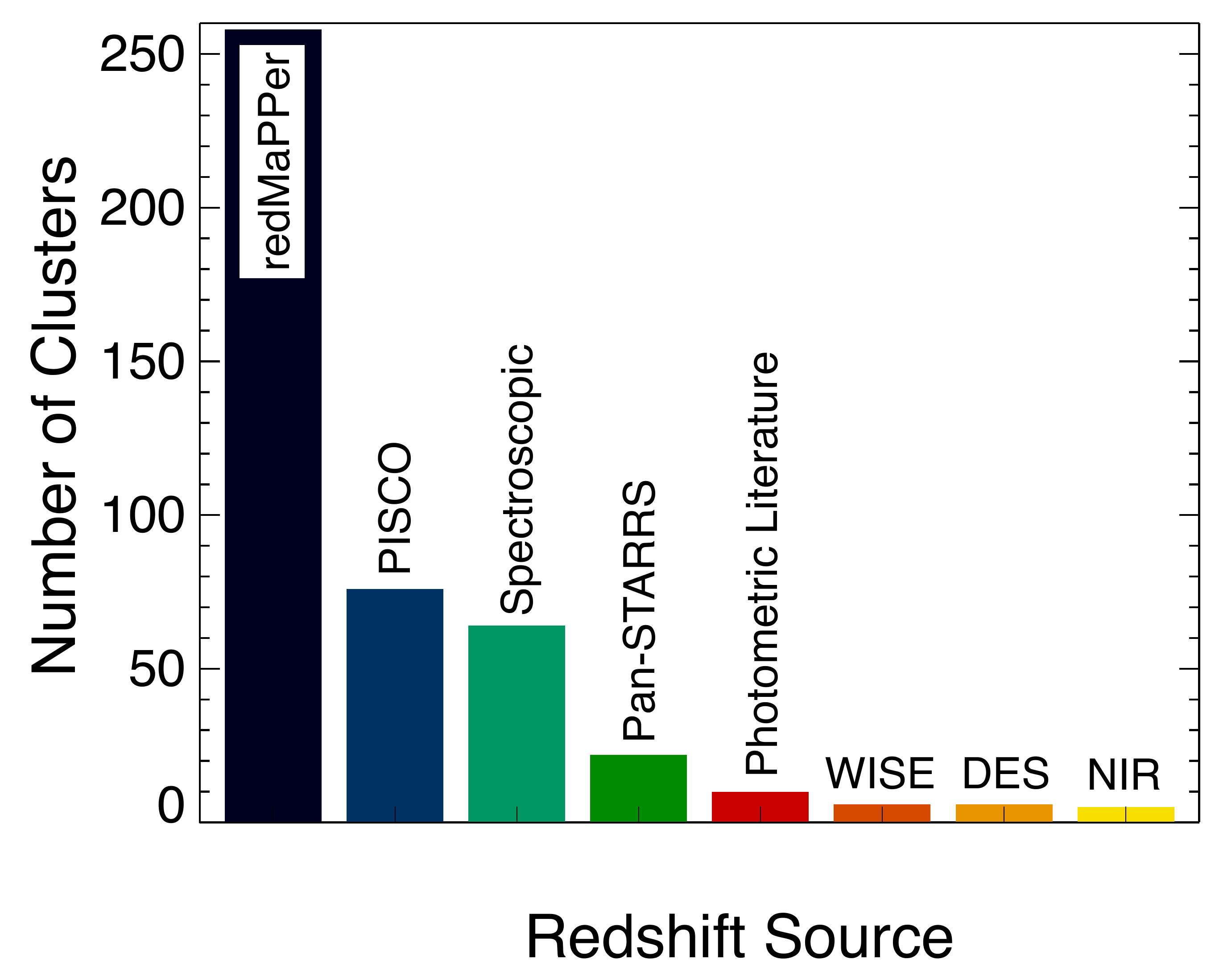}
\caption{Distribution of the telescope/surveys used to confirm and provide redshifts for the SPT clusters presented in this catalog. While some clusters may have redshifts from multiple sources (for example there is significant overlap between PISCO and RM), we only represent each cluster once in this figure, highlighting the sources of the redshifts reported in Tables \ref{tab:sampletable} and \ref{tab:sample4sigma}. The DES column corresponds to clusters with redshifts from DES data but not from RM (see Section \ref{subsec:otherz}). Generally, data from  Pan-STARRS  is deep enough to confirm clusters to $z\sim0.6$, DES and PISCO to $z\sim0.8-1.0$,  FourStar to $z\sim1.2$ and  \textit{Spitzer}  to $z\sim1.5$. 
}
\label{fig:redshift-source}
\end{figure}

We adopt two different techniques for confirming candidates in clusters: a probabilistic matching to RM clusters in the common overlap region and, for candidates outside the volume probed by RM clusterfinding on DES data, the identification of significant over-densities of red sequence or 1.6 $\mu$m rest-frame galaxies at the locations of cluster candidates using the techniques described in B15. 
We show the distribution of the origins of \surveyshort \ cluster redshifts in Figure~\ref{fig:redshift-source}. 

\subsubsection{Confirmation with redMaPPer in Scanning Mode}\label{subsec:desmatch}

\begin{figure}[th] 
\hspace*{-0.25in}
\includegraphics[width=3.5in]{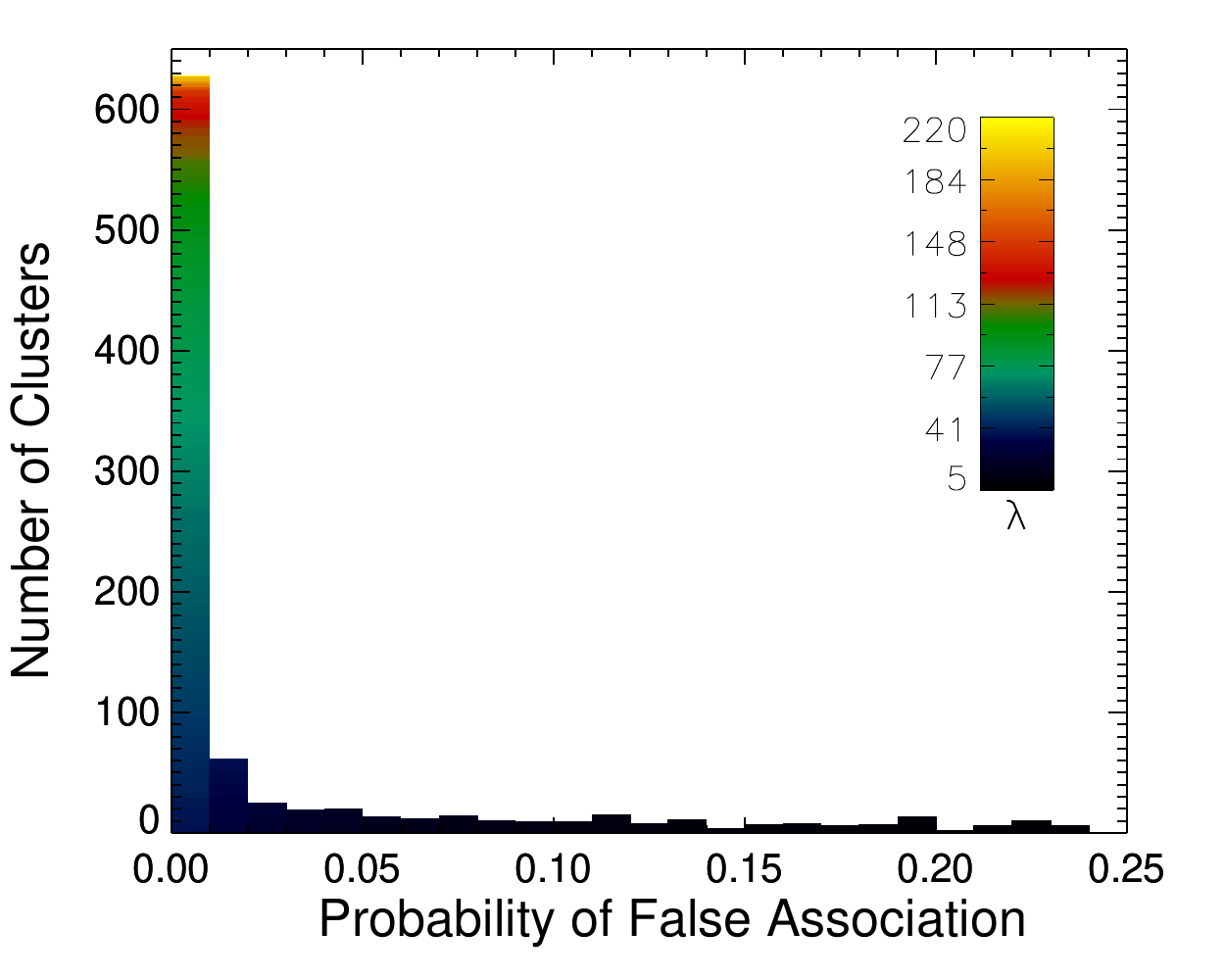}
\caption{Distribution of probabilities of false association between redMaPPer targeted confirmations and SPT clusters from the \surveyshort \ and SPT-SZ surveys. The color scaling represents the optical richnesses of the RM detections centered on SPT locations. For the purposes of this work we reject associations with probabilities of false association greater than 5\%.} 
\label{fig:des_matches}
\end{figure}

The RM catalog makes strict cuts on sky coverage and photometric depth to ensure a well-understood optical selection function. 
In the case of matching to an SZ-selected sample we can somewhat relax these criteria to also enable targeted searches for red-sequence galaxy counterparts in regions excluded by these cuts. 
We have run the RM algorithm in ``scanning mode''  centered on the SPT location where the likelihood of a red-sequence overdensity in apertures of 500 kpc radius is computed as a function of redshift from $z$=0.1 to 0.95 in steps of $\delta_z=0.005$.   
At each redshift the optical richness is computed at both the SZ location and the most likely optical center; for systems with significant red-sequence overdensities the richness and redshift is refined at the highest likelihood redshift using the standard RM radius/richness scaling.  Richnesses are recorded for each location where $\lambda \ge 5$. We repeat this scanning procedure for 100 mock SZ samples (at over 100,000 sky locations) to compute the probability over random of finding a cluster of richness $\lambda$ at an SZ location. 
We report as ``confirmed'' clusters for which the probability of random association is less than 5\%, which corresponds to $\lambda>19.3$.  As this probability distribution is a continuum (with no clear breaks) this choice is somewhat arbitrary; setting this threshold at 0.05 leads to an expectation of $< 2$ false associations in the RM-confirmed sample.
In Figure \ref{fig:des_matches} we plot the distribution of matched clusters against the probability of random associations for SPT-SZ ($\xi>4.5$) and \surveyshort \ clusters ($\xi>4$).

For cluster candidates in the common region not confirmed via the RM scanning-mode process we make use of both DES and WISE imaging and photometric catalogs at the cluster locations 
and, where available, pointed follow-up imaging as described in the next subsection. 
The confirmation of these clusters follows a similar process to that described below.

\subsubsection{Cluster confirmation from other imaging datasets}\label{subsec:otherz}

\begin{figure}[t]
\hspace*{-0.4in}
\includegraphics[width=3.7in]{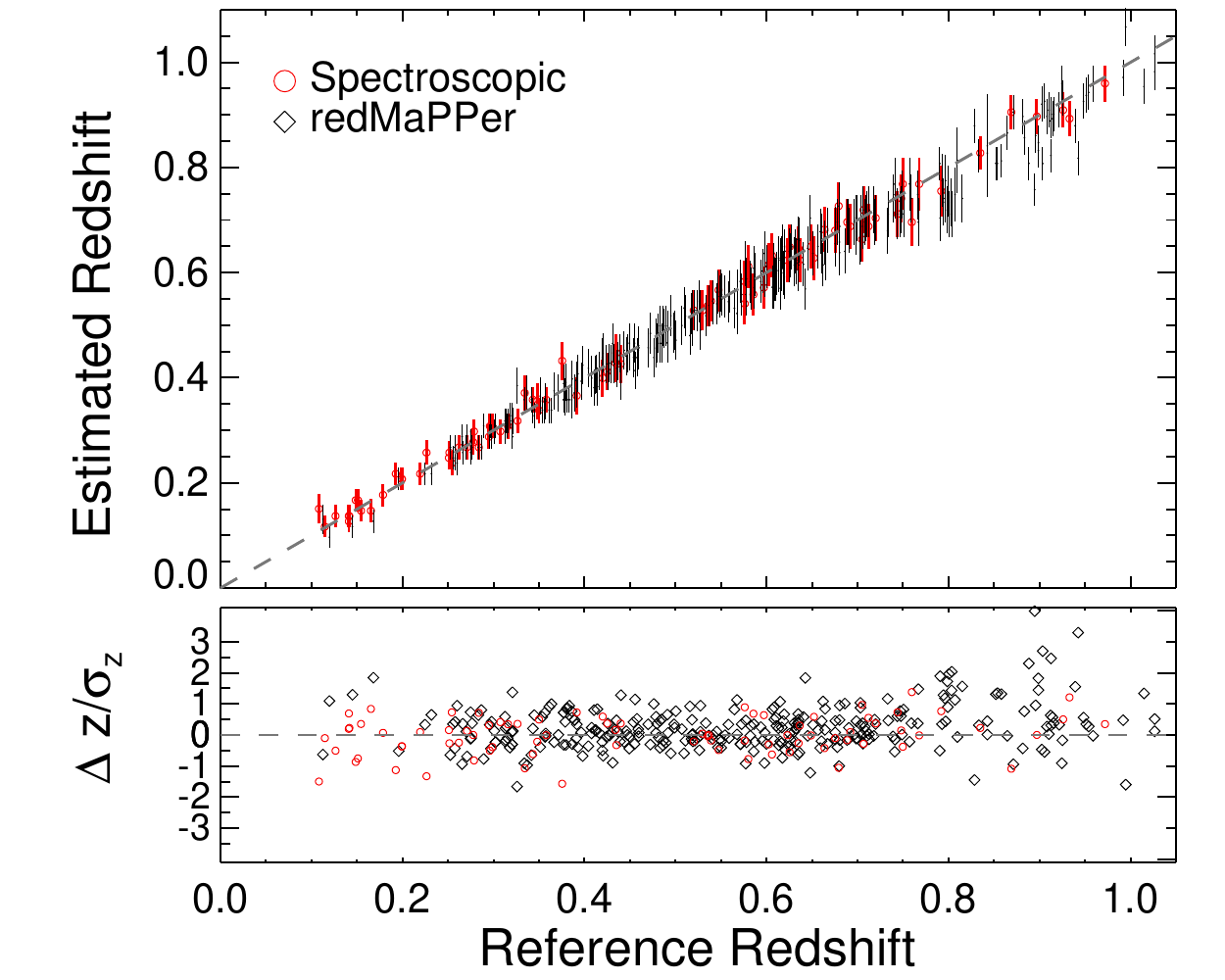}
\caption{(Top panel) Comparison of estimated red-sequence redshifts from PISCO imaging data to spectroscopically measured redshifts (81 systems; red) and redshifts estimated by the redMaPPer scanning-mode algorithm (318 systems; black). (Bottom panel) The distribution of residuals over the redshift uncertainties; for the RM-PISCO comparison we add the individual uncertainties in quadrature.  In general we find excellent agreement between the redshifts measured from PISCO and  both the RM and spectroscopic samples below $z\sim0.75$. }
\label{fig:redshift_tuning}
\end{figure}

Here we describe the techniques used for confirming cluster candidates not confirmed via the RM algorithm or literature search.  We obtained imaging for $\sim 100$ candidates outside of the volume searched by RM as well as some imaging redundant to the DES imaging (in terms of confirmation) as part of our strong lensing search program that we use here for redshift comparisons.   
In total, 173 candidates were imaged with Magellan/PISCO (about 2/3 in common with RM, see Figure Figure \ref{fig:redshift_tuning}), 19 with Magellan/Fourstar, and 7 with Spitzer (note the NIR imaging overlaps areas with optical imaging). 10 candidates are located in the DES footprint but are either at high redshift or are missing photometry in filters required for RM, and 22 candidates only fall in the Pan-STARRS footprint.
 
To conduct our targeted search for red-sequence galaxies in these areas, we first calibrate our synthetic model for the colors and magnitudes of red-sequence galaxies, generated with the {\tt GALAXEV} package \citep{bruzual03} by assuming a passively evolving stellar  population with single formation burst at $z=3$, to match the relevant survey photometry using samples of known clusters with spectroscopic redshifts. 
For PISCO, the dataset that we most use to confirm clusters outside of DES, we use 58 SPT-SZ galaxy clusters with spectroscopic redshifts that were imaged as part of our broader SPT characterization program. 
In Figure \ref{fig:redshift_tuning} we plot in red the measured PISCO redshifts versus those from the training sample as well as additional \surveyshort \ clusters with spectroscopic redshifts reported in the literature. 
The typical redshift precision is $\sigma_z/(1+z)\sim 0.015$ with uncertainties increasing towards higher redshifts. 
We also plot in black the PISCO redshifts compared to those from the DES RM catalog for 318 systems in SPT-SZ and \surveyshort \ and find generally good agreement between the two, though the comparison suggests that the redshifts estimated from PISCO may tend be underestimated at the highest redshifts. More spectroscopic data on high-redshift clusters from ongoing SPT programs will help further validate/improve the PISCO redshift calibration for such systems.  
Given the excellent redshift precision of the RM algorithm, we adopt RM scanning-mode redshifts by default when clusters are confirmed by both methods.
We repeat a similar process with DES photometry, finding $\sigma_z/(1+z) \sim 0.015$, and with 35 spectroscopic clusters (as identified in the \surveyshort \ literature search)  at $0.108 < z < 0.72$ in the Pan-STARRS1 footprint, finding $\sigma_z/(1+z) \sim 0.03$ in these shallower data.

We search the optical/NIR imaging for an excess of red-sequence galaxies (or alternatively 1.6 $\mu$m rest-frame galaxies in the case of \textit{Spitzer} and WISE confirmations) in the vicinity (2-3\arcmin) of the SPT cluster candidates.
We call a cluster ``confirmed'' when significant excesses of these galaxies over background are identified  (see e.g., B15 for more details on the confirmation procedure). 
In \citet{song12b} we estimated that $<4\%$ of cluster candidates identified via this procedure would be false associations and, for clusters in common between the PISCO and DES imaging, we can cross check our assigned confirmations against the statistical process described in Section \ref{sec:des}. 
We note that this is of course a lower limit to the false association rate as the DES data is also of finite redshift reach. 
In this comparison we find that $\sim1\%$ (3/318) of candidates with RM cluster matches that were also targeted with PISCO were assigned a different cluster counterpart when using the PISCO data. 
In two circumstances the PISCO data were insufficiently deep to correctly confirm the higher-redshift ($z\sim0.9$)  clusters while the remaining system was a superposition of two rich clusters ($\lambda=75$ and $\lambda=65$) for which the targeted RM algorithm selected the lower redshift system as the richer cluster and the PISCO data the higher. 

For confirming higher-redshift systems without targeted \textit{Spitzer} or Magellan/Fourstar data we combine data from ``unWISE'' with optical source catalogs. 
Following \citet{gonzalez19}, we adopt a 1\farcs5 \ matching radius to associate WISE sources with optical galaxies and exclude sources with $i<21.3$ and W1-W2 $<0.2$ as these cuts were found to remove low-redshift ($z<0.8$) galaxies. 
Similar to the analysis of clusters with \textit{Spitzer} imaging, we search for a local excess of galaxies at 1.6~$\mu$m rest frame in the vicinity of the SPT cluster candidates. 
We validated this search process on clusters from the SPT-SZ sample (B15; \citealt{khullar19}) with spectroscopic redshifts $z>0.85$, finding that we were able to robustly confirm 20/23 of these systems. 
From this spectroscopic sample we were able to quantify the redshift uncertainty in our WISE measurements as $\sigma_z/(1+z)\sim 0.1$. 
Improving this redshift precision via more sophisticated catalog cuts and photometric analysis of the WISE data is work in progress.

\subsubsection{2dFLenS}\label{sec:tdfsec}
The 2dFLenS survey  \citep{blake16} targeted luminous red galaxies (LRGs) at $z<0.9$ with a primary focus on measuring redshift-space distortions and---in combination with KiDS survey data---galaxy-galaxy lensing \citep{joudaki18} and the characterization of redshift distributions via cross-correlation \citep{johnson17}. 
There is significant overlap between the southern field of 2dFLenS and \surveyshort. 
A number of  visually identified brightest cluster galaxies from SPT clusters were targeted in a spare-fiber program (though all but two were lost owing to weather) and here we identify additional 2dFLenS sources associated with SPT clusters. 
First, for each confirmed candidate in the SPT sample at $z<0.9$, we search for spectroscopic LRGs within 2.5\arcmin \ of the cluster location and find 47 systems with spectroscopic galaxy associations. 
Repeating the process on the 40 random position catalogs provided by the 2dFLenS team\footnote{http://2dFLens.swin.edu.au/} we find an average of 17 such matches per mock catalog, resulting in $\sim$30 matches over random for the real data sample.
We further improve the purity of the matching by restricting matched galaxies to have redshifts within 2$\sigma$ of the photometric redshift error (or $\delta z < 0.015$ for clusters with spectroscopic redshifts) and find 39 clusters with spectroscopic galaxy counterparts including 2 systems that were targeted as part of the spare fiber program, compared to an average of 2 systems for the random catalogs.  We list these systems in the Appendix in Table~\ref{tab:tdf_match}.

\section{The Cluster Sample}

\begin{figure*}[t]
\includegraphics[width=7in]{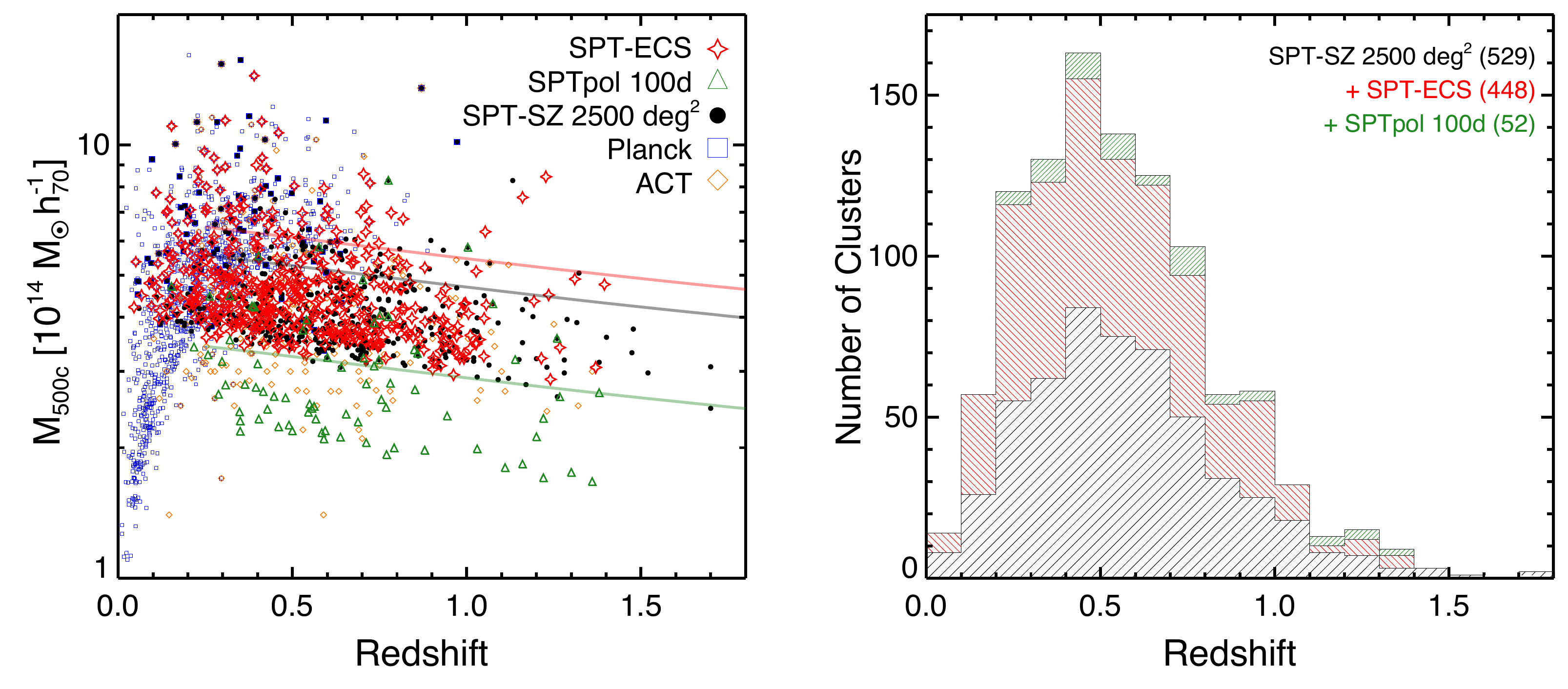}
\caption{(Left) The mass and redshift distribution of the \surveyshort \ cluster sample detected at $\xi \ge 4$. The median redshift of the sample is $z=0.49$ and the median mass is $M_{500c} \sim {4.4 \times 10^{14} M_\odot h^{-1}}$.  Overplotted are cluster samples from other SZ surveys including  the 100d SPTpol survey (green triangles; \citealt{huang19}), the 2500d SPT-SZ Survey (black circles; \citealt{bleem15b}, with redshifts updated as in \citealt{bocquet19}); the PSZ2 cluster sample from \citet{planck15-27} (blue squares), and the cluster samples from the ACT collaboration (orange diamonds; \citealt{hasselfield13, hilton18}). Clusters found in both SPT and other samples are plotted at the SPT mass and redshift and, for clusters in common between other samples, at the mass and redshift at which the cluster was first reported. We also plot at $z>0.25$, as solid colored lines, the 90\% completeness thresholds for $\xi \ge5$ for the three SPT surveys (see also Figure \ref{fig:completeness}). (Right) A redshift histogram of the three reported SPT cluster surveys. The number of clusters in each survey---with each cluster only reported once (so that e.g., clusters in both SPTpol 100d and SPT-SZ are only counted once)---are listed to the right of each survey name. The contribution from the SPTpol 100d survey is plotted on top in green right-diagonal hatch, the contribution from the SPT-ECS survey is plotted in red left-diagonal hatch, and the contribution to the total from the SPT-SZ survey is plotted in black right-diagonal hatch. Combined with these other two samples, the SPT-ECS sample brings the number of SZ-detected clusters reported by the SPT collaboration to over 1,000.}
\label{fig:mass_redshift}
\end{figure*}

In this section we describe the new SZ-selected cluster sample. We also compare the properties of these clusters to those of SZ-selected clusters identified by \textit{Planck} in the \surveyshort \ region.
Using the confirmation criteria presented in Section  \ref{sec:confirm}, we confirm  \nconfirmfive \ of \ncandfive \ candidates at $\xi \ge 5$. 
We also leverage the DES and other imaging data to confirm an additional \nconfirmfourtofive \ clusters at $4<\xi<5$ but note that while the DES imaging is sufficient for cluster confirmation out to $z\sim0.8-1.0$ in the \surveyshort-DES overlap region, our follow-up of this lower-significance sample is otherwise highly incomplete. 

While the confirmation process is still ongoing, we can compare these numbers to our expected numbers of false detections as estimated in Section \ref{subsec:false}. 
As discussed in B15, expectations from simulations were found to be in good agreement with observations of the more uniformly and deeply imaged SPT-SZ sample. 
At  $\xi \ge 5$ where our optical follow-up imaging is sufficient to confirm clusters to at least  $z\sim0.85$, we find 22 unconfirmed candidates compared to the expected $21\pm 4$. 
\textcolor{black}{This places an empirical lower limit on the purity of 91\% for the $\xi > 5$  SZ candidate sample which, when compared to the simulation prediction, suggests that there are relatively few clusters that remain to be confirmed.
For  the $\xi \ge 4.5$ SZ candidate sample, where the follow-up is generally more heterogeneous/incomplete,  we find 
180 currently unconfirmed candidates compared to $174\pm13$ expected, resulting in a lower limit to the purity of  $64\%$.}

The confirmed cluster candidates have a median redshift of {$z=0.49$} and median mass (calculated as described below in Section \ref{subsec:szmass}) of $M_{500c} \sim {4.4 \times 10^{14} M_\odot h^{-1}}$.  
Twenty-one of the systems are at $z>1$, bringing the total number of $z>1$ systems from SPT-SZ, SPTpol 100d \citep{huang19}, and \surveyshort \ to over 75 out of  $>1,000$ confirmed systems. 
The mass and redshift distribution of the cluster sample as compared to other SZ-selected samples, as well as a histogram of the redshift distribution of the SPT samples, are shown in Figure \ref{fig:mass_redshift}.  
We note that, given the lack of deep NIR data redder than $z-$band, the RM algorithm can systematically underestimate redshifts at $z>0.9$ which may be the source of the small gap in the cluster redshift distribution at $z\sim1.1$. 

\begin{figure}
\includegraphics[width=3.5in]{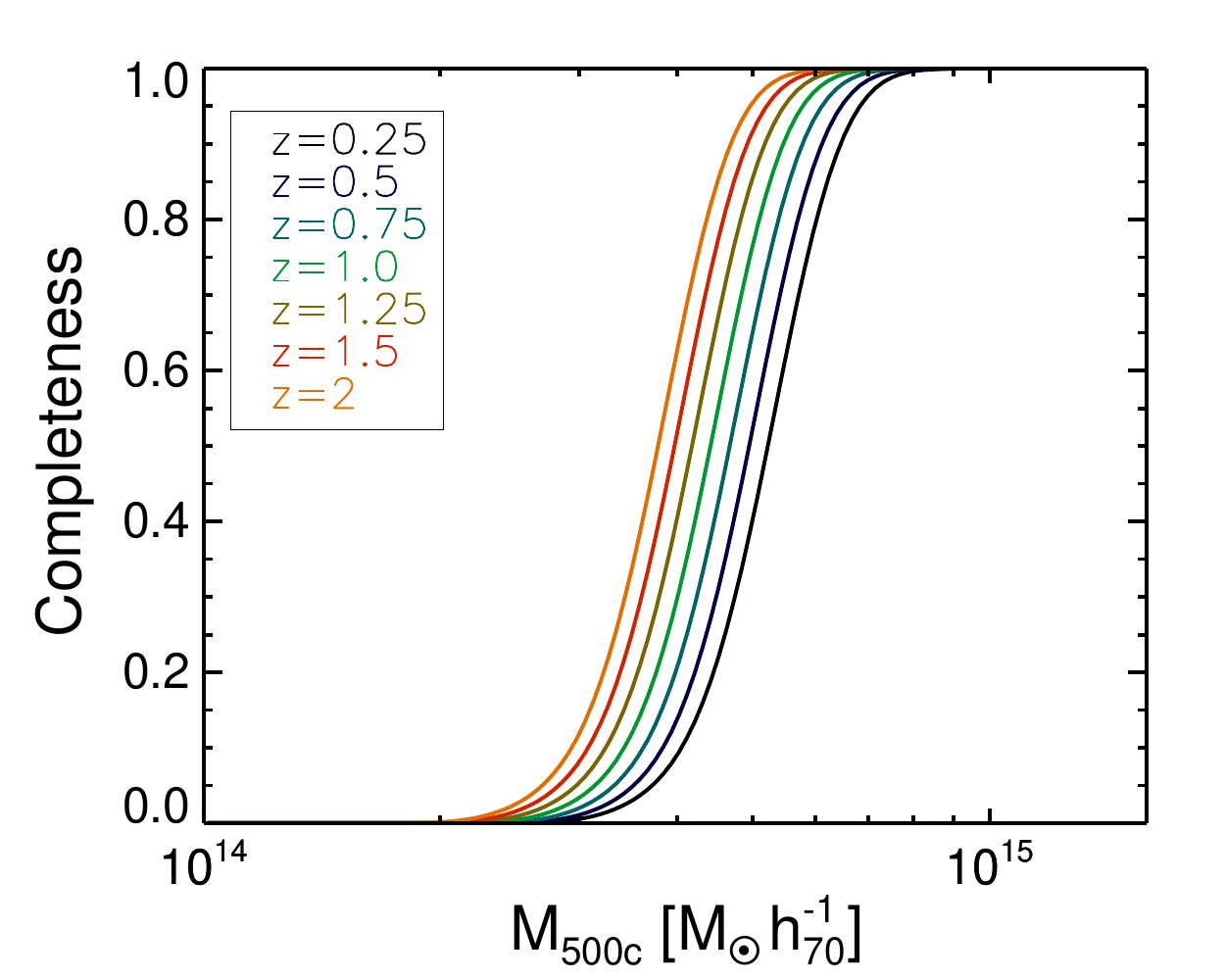}
\caption{The average (across all fields) completeness as a function of mass and redshift for the \surveyshort \ survey at $\xi\ge5$, the $\xi$ threshold for which we provide the complete candidate list for the full survey. The survey is on average 90\% complete for all redshifts at $z>0.25$ for masses above $M_{500c} \sim {6.5 \times 10^{14} M_\odot h^{-1}}$. 
This completeness is derived for a fixed cosmology as discussed in Section \ref{subsec:szmass} and \citet{bocquet19}.}
\label{fig:completeness}
\end{figure}

In  Figure \ref{fig:completeness}, we present  an estimate of the survey completeness as a function of mass  and redshift for our main sample at $\xi>5$ using the $\xi-$mass relation (see below in Section \ref{subsec:mass}). 
The survey is on average $>90\%$ complete at all redshifts $z>0.25$ for masses above $M_{500c} \sim {6.5 \times 10^{14} M_\odot h^{-1}}$ (in comparison to $M_{500c} \sim {5.5\times 10^{14} M_\odot h^{-1}}$ for the SPT-SZ survey at the same significance threshold), with the mass at which the survey is 90\% complete  shifting by less than $1 \times 10^{14} M_\odot h^{-1}$ from the mean between the fields.  
The mass corresponding to a fixed completeness value falls as a function of redshift, with the survey on average 90\% complete at  $M_{500c} = {5.4 \times 10^{14} M_\odot h^{-1}}$ at  $z>1$.
In Table \ref{tab:sampletable}, we provide a complete listing of the candidates at $\xi \ge 5$ including their positions,  detection significances and the filter scales that maximize these significances.  
For confirmed clusters we also include redshifts, estimated masses, optical richness measures (where available), and we flag notable properties about the systems.  
In Table \ref{tab:sample4sigma} we provide a similar listing for the lower-significance confirmed systems. 

As mentioned in Section \ref{subsec:lit} we also conducted a literature search for previously identified clusters, finding 147 \surveyshort \ candidates have been previously reported including a number of systems in the Abell and \textit{Planck} cluster samples \citep{abell58,abell89,planck15-27} as well small numbers of systems in other samples (e.g., APM, MACS, SWXCS, MaDCoWS, \citealt{dalton97,cavagnolo08,mann12,liu15b,gonzalez19}).
By far the largest overlap is with the \textit{Planck} PSZ2 sample; we explore this in more detail in Section \ref{subsec:planckc}.

\subsection{Comparison to the SPT-SZ Cluster Abundance}
\label{subsec:validateNC}

The cluster catalog extracted from \surveyshort \ should be statistically consistent with
the catalog extracted from the SPT-SZ survey once the different survey
properties such as depth are accounted for. To test this, we use a cluster number
count (NC) analysis to calibrate the parameters of the $\xi$--mass scaling
relation assuming a fixed cosmology and compare the results with those obtained
for SPT-SZ.

\subsubsection{The SZ $\xi$--Mass Relation}\label{subsec:szmass}

To connect the observed SZ significance, $\xi$, to cluster mass we adopt an observable--mass scaling relation of the form 
\begin{equation}
\langle\ln\zeta\rangle = \ln\bigl[A_\textrm{SZ} \left( \frac{\mass}{3 \times 10^{14} M_{\odot} h^{-1}} \right)^{B_\textrm{SZ}} \left(\frac{H(z)}{H(0.6)}\right)^{C_\textrm{SZ}}\bigr],
\label{eqn:zetam}
\end{equation}
\begin{equation}
P(\ln\zeta|M,z) = \mathcal N\left[\langle\ln\zeta\rangle(M,z), \sigma_{\ln\zeta}^2\right]
\end{equation}
where $A_\textrm{SZ}$ is the normalization, $B_\textrm{SZ}$ the slope, $C_\textrm{SZ}$ the redshift evolution, $\sigma_{\ln\zeta}$ the log-normal scatter on $\zeta$, and $H(z)$ is the Hubble parameter.
The variable $\zeta$ represents the ``unbiased"  significance that accounts for the maximization of $\xi$ over position and filter scales during cluster detection:
\begin{equation}
\label{eq:zetaxi}
P(\xi|\zeta) = \mathcal N(\sqrt{\zeta^2+3}, 1)
\end{equation}
for $\zeta > 2$ \citep{vanderlinde10}. 
As in previous SPT publications, we rescale $A_\textrm{SZ}$ on a field-by-field basis to account for the variable depth of the survey: $A_\textrm{SZ,field} = \gamma_\textrm{field} \times A_\textrm{SZ}$ (e.g., \citealt{reichardt13, dehaan16}).
These field renormalization factors, $\gamma_\textrm{field}$, are computed using the simulations described in Section \ref{subsec:false} and are reported in Table \ref{tab:fields} on the same reference scale as the analogous factors for the SPT-SZ survey.

The different fields of the SPT-SZ and \surveyshort \ surveys have a small amount of
overlap  at the field boundaries. We correct the field areas such that the total
effective survey  area corresponds to the unique sky area that is surveyed.
These corrections are between $0.03\%$ and $1.9\%$. SZ detections in the field
overlap regions are matched by keeping the candidate with the larger detection
significance $\xi$. Note that this approach is different from the one adopted in
\cite{dehaan16, bocquet19}, who double-counted clusters in the field overlap
regions in SPT-SZ in their NC analyses. The exact treatment of the field boundaries has negligible
impact on our results; for example, the change in our total predicted cluster counts due to not
correcting for the field overlap area is much smaller than the recovered
uncertainty.

\subsubsection{$\gamma_\mathrm{ECS}$ Constraints from the Cluster Abundance} \label{sec:abundance}

We model the cluster sample as independent Poisson draws from the halo mass
function. The likelihood function for the vector of cosmological and scaling relation
parameters $\vec p$ is
\begin{equation} \label{eq:like_NC}
  \begin{split}
    \ln \mathcal L(\vec p) =  & \sum_i \ln \frac{dN(\xi_i, z_i| \vec p)}{d\xi dz} \\
    &- \int_{z_\mathrm{cut}}^\infty dz \int_{\xi_\mathrm{cut}}^\infty d\xi \frac{dN(\xi, z| \vec p)}{d\xi dz}.
  \end{split}
\end{equation}
The sum runs over all clusters $i$ in our sample, and
\begin{equation}
\begin{split}
\frac{dN(\xi, z |\, \vec p)}{d\xi dz} =
\iint & dM\, d\zeta\, \bigl[ P(\xi|\zeta)P(\zeta|M,z,\vec p)\\
&\frac{dN(M, z| \vec p)}{dM dz}\Omega(z,\vec p) \bigr]
\end{split}
\end{equation}
where $\Omega(z,\vec p)$ is the survey volume and $dN/dM dz$ is
the halo mass function given by \citet{tinker08}.
The second line in Eq.~\ref{eq:like_NC} corresponds to the total number of clusters in the survey.

We analyze the \surveyshort \ NC assuming our fixed $\Lambda$CDM cosmology.
By construction of our scaling relation model, the amplitude $A_\mathrm{SZ}$ and
the correction factor $\gamma_\mathrm{ECS}$ (introduced in Section \ref{subsec:false})
are fully degenerate. The constraints on the SZ scaling relation
parameters $B_\mathrm{SZ}$, $C_\mathrm{SZ}$, and the scatter
$\sigma_{\ln\zeta}$ from SPT-SZ and \surveyshort\ are consistent at the $\ll1\sigma$ level.
To test the consistency of the relative scaling between the two surveys,
we analyze the joint NC from SPT-SZ and \surveyshort.
In this analysis, any residual in the relative calibration between the two surveys is absorbed by $\gamma_\mathrm{ECS}$.
We recover
\begin{equation}
  \gamma_\mathrm{ECS} = 1.124 \pm 0.045.
\end{equation}
We provide and discuss an alternate calibration of $\gamma_\mathrm{ECS}$ in section~\ref{subsec:lambdam}.

\subsection{Mass Estimation}\label{subsec:mass}

The $\xi$--mass relation defined above in Eqs.~\ref{eqn:zetam}--\ref{eq:zetaxi} allows us to compute mass estimates for all sample clusters.
We adopt $A_\textrm{SZ}=4.08$, $B_\textrm{SZ}=1.65$, $C_\textrm{SZ}=0.64$, and $\sigma_{\ln\zeta}=0.20$.
These mean scaling relation parameters were determined in \citet{bocquet19} for our fixed reference flat \lcdm{} cosmology 
and using the SPT-SZ sample at $\xi>5$ and $z>0.25$. As discussed in the previous section, the SZ scaling relation parameters barely shift between
 a NC analysis using SPT-SZ clusters alone and one using SPT-SZ and SPT-ECS clusters,
and we thus use the SPT-SZ only numbers presented in \cite{bocquet19} for consistency with their mass estimates.

\subsection{Comparisons to the Planck Cluster Sample}\label{subsec:planckc}

There is naturally significant overlap between the \textit{Planck} and \surveyshort \ cluster samples as both identify massive clusters by the SZ effect. 
Here we focus our comparison on the reported masses and redshifts, two quantities critical for cosmological analyses.
To directly compare the properties of the two samples for clusters in common we first associate the catalog from \citet{planck15-27}  with the \surveyshort \ catalog using a $4\arcmin$ matching radius  
and find that 82 SPT candidates (81 confirmed clusters) detected at  $\xi>4$ match \textit{Planck} systems within this radius.

Overall we find good agreement between the redshifts for matched clusters, with three outliers for which the estimated redshifts reported in the \textit{Planck} catalog and this work differ by $\delta_z > 0.1$.  
These three systems each have redshifts in this work from the RM algorithm.  Two of the systems (J0046$-$3911 and J0516$-$2236) have photometric redshifts reported in the literature while 
the third system, J0348$-$2144  (PSZ2~G215.19$-$49.65, separated from the SPT position by 0\farcm58), was associated in the  \textit{Planck} catalog with ACO 3168 (RXC~J0347.4$-$2149) for which a spectroscopic redshift of $z=0.2399$ was derived from 5 cluster members in \citet{chon12}.
This system is significantly offset (8\farcm6, 8\farcm9) from the SPT and \textit{Planck} detections, respectively. 
 We instead associate this cluster candidate with a closer (1\farcm2, 1\farcm7)  and richer  ($\lambda=186$ versus 10) system at $z=0.347 \pm 0.008$. 

Beyond the direct redshift comparisons, we also provide here redshifts for 13 PSZ2 systems from \citet{planck15-27}; 11 of these clusters were not confirmed by the Planck collaboration. These systems are listed in Table \ref{tab:confirmed_planck}. 
Two of these clusters have previously reported redshifts in \citet{maturi19} and a third we associate with ACO S 1048 \citep{abell89}. We find good agreement with the \citet{maturi19} redshift estimate for PSZ2~G011.92$-$63.53 but find $\delta_z > 0.2$ for PSZ2~G011.36$-$72.93.

\begin{figure}
\includegraphics[width=3.5in]{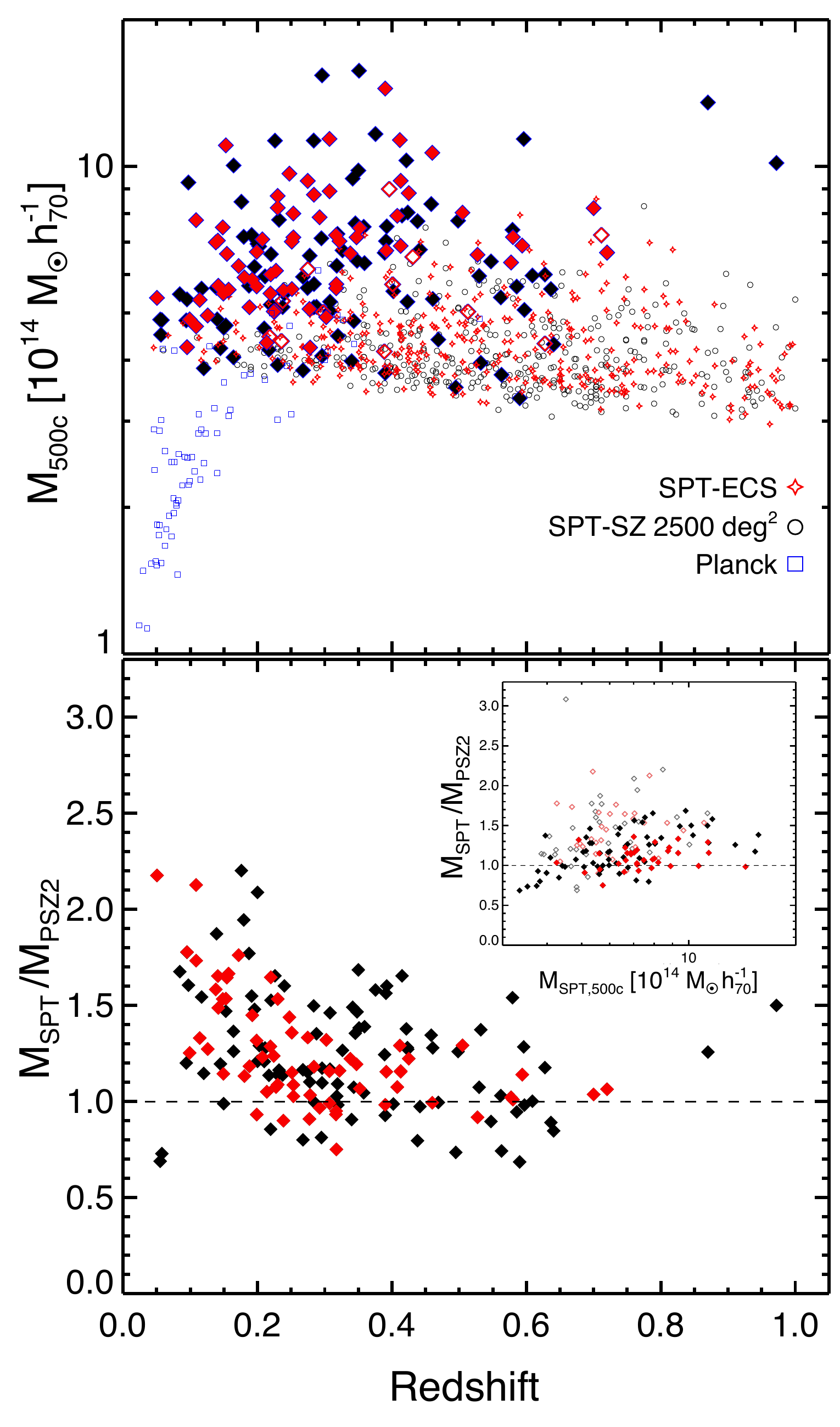}
\caption{Top panel: Mass versus redshift for SPT and \textit{Planck} clusters in the sky area surveyed by SPT. Small open symbols represent clusters that reside only in one of the catalogs while the filled diamonds represent clusters (plotted at the  SPT masses) that are in both SPT-SZ and \textit{Planck} (black, blue outline) and \surveyshort \ and \textit{Planck}  (red, blue outline). The 13 \textit{Planck} clusters for which we report a redshift in Table  \ref{tab:confirmed_planck} are plotted as large hollow diamonds. In this plot we omit \textit{Planck} clusters that fall within regions excluded by the SPT point source veto (see Section \ref{sec:masking}) and restrict the x-axis to focus on redshifts where the samples overlap. Bottom panel: Again using red diamonds for clusters in common between \textit{Planck} and SPT-ECS and black diamonds for those in \textit{Planck} and SPT-SZ, we plot the ratio of reported SPT to \textit{Planck} masses versus redshift and (inset) versus SPT mass. In the inset panel clusters at $z\ge0.25$ $(z<0.25)$ are plotted as filled (open) symbols. }
\label{fig:planckcompare}
\end{figure}

We can also compare the reported SZ mass estimates for each of these samples. 
In Figure \ref{fig:planckcompare} we show the \textit{Planck} and SPT clusters in the \surveyshort \ and SPT-SZ footprints. In the top panel, plotted as small hollow symbols, are clusters that are only found in one of the catalogs, while the filled diamonds represent clusters that are in both SPT and \textit{Planck}. 
The 13 \textit{Planck} clusters for which we report a redshift from \surveyshort \ in Table  \ref{tab:confirmed_planck} are plotted as large hollow diamonds. 
Including clusters from the SPT-SZ region brings the joint SPT-\textit{Planck} sample to a total of 150 clusters with mass estimates for which the reported redshifts differ by $\delta_z < 0.1$\footnote{We implement the redshift cut so that the masses would not be significantly different simply from the use of different redshifts in the mass estimation process.}, and 88 such systems at $z>0.25$ where SPT masses are expected to be unbiased.  
In the bottom panel we plot the ratio of SPT to \textit{Planck} mass as a function of redshift and, in the inset, as a function of the SPT mass estimate. Qualitatively we notice a trend with redshift where 
at $z<0.25$ the ratio of the SPT to \textit{Planck} masses is significantly higher than at higher redshifts (1.44$^{+0.05}_{-0.14}$ vs. 1.1$^{+0.055}_{-0.03}$). 
We note that mass estimates for SPT clusters at $z<0.25$ are more uncertain---though not expected to be biased high---given increased noise contributions from both the primary CMB and atmosphere as well as the removal of large scale sky signal by the map filtering.

Comparisons to the \textit{Planck} SZ masses are often reported in terms of a mass bias, {1-\textit{b}}, where {$M_\textit{Planck} = (1-b)M_\textrm{True}$}. For purposes of comparison here we treat the SPT masses as the ``true'' cluster masses and both compute the median mass bias and check for a mass-dependent trend.
The latter is achieved via making use of the Bayesian linear regression routines provided by  \citet{kelly07}, and fitting for the power-law index, $\alpha$:
\begin{equation}
M_\textrm{500c Planck} \propto M_\textrm{500c SPT}^{\alpha}.
\end{equation}
Here we consider only the statistical errors in the SPT and \textit{Planck} masses as we are directly comparing properties of the same clusters. 

As discussed in \citet{battaglia16}, one must take care in such comparisons as they can be impacted at the level of 3-15\% in {(1-\textit{b})}  by the presence of Eddington bias in the reported \textit{Planck} masses.\footnote{See e.g., \url{https://wiki.cosmos.esa.int/planckpla2015/index.php/Catalogues\#SZ\_Catalogue}}  
We follow \citet{battaglia16} and recompute the SPT masses not accounting for this bias.
Restricting to a subset of 69 clusters where the absolute difference between the \textit{Planck} and SPT signal-to-noise was less than two (so the bias would be somewhat comparable), 
we find a median (1-\textit{b})$_\textrm{Eddington}$ = $0.77^{+0.02}_{-0.025}$  and $\alpha=1.03\pm0.14$ for the full sample and (1-\textit{b})$_\textrm{Eddington}$ = $0.80^{+0.09}_{-0.01}$ and $\alpha=1.3 \pm 0.27$ for 15 such clusters with $0.25<z<0.35$ and uncorrected $M_{500c \textrm{ SPT}} > 5.5\times10^{14}$ (where both samples are more complete). 
To aid comparison with previous studies in the literature we also compute these values for the entire matched sample with \textit{debiased} SPT masses, finding (1-\textit{b}) = $0.83\pm0.02$ for the full matched sample and (1- \textit{b})= $0.91^{+0.01}_{-0.05}$ and $\alpha=0.75 \pm 0.06$ at $z>0.25$.

Comparisons between (Eddington-biased) \textit{Planck} and (debiased) SPT mass estimates were previously reported by \citet{planck15-27} and \citet{hilton18} for the SPT-SZ and PSZ2 samples. 
\citet{planck15-27} found the SPT-reported masses to be on average 20\% higher than the \textit{Planck} masses---in good agreement with the results derived above with the larger SPT-SZ and \surveyshort \ sample. 
\citet{hilton18} explored the relation between SPT-SZ, ACT, and PSZ2  masses finding the mean mass ratio of ACT to SPT clusters to be $1.00\pm0.04$  for 18 clusters in common between the samples; the \surveyshort \ sample provides no additional overlapping systems between ACT and SPT to further this comparison. \citet{hilton18} additionally noted a mass-dependent trend between the \textit{Planck} and SPT/ACT masses, finding for the ACT comparison $\alpha=0.55\pm0.18$, a result $\sim1\sigma$ lower than our value. 

A number of studies have also contrasted the estimated \textit{Planck} masses against masses estimated using other observables, with values of (1-\textit{b}) ranging from $\sim0.7$ to unity 
(e.g.,  \citealt{vonderlinden14b} $, 0.688\pm0.072$; \citealt{hoekstra15}, $0.76\pm0.05$; \citealt{smith16}, $0.95\pm0.04$; \citealt{medezinski18}, $0.80\pm0.14$).  
Our recovered values fall within this range.  
Other works report values for the power-law index, $\alpha$ (e.g., \citealt{schellenberger17}, $0.76 \pm 0.08$;  \citealt{mantz16}, $0.73 \pm 0.02$) consistent with our measurement when using debiased SPT masses.

\textcolor{black}{We also examine the  \surveyshort \ footprint for clusters detected by \textit{Planck} but not by SPT. }
Based on the selection function shown in Figure \ref{fig:completeness}, we expect the \surveyshort \ sample to contain  essentially all confirmed \textit{Planck} clusters at $z\ge0.25$ in the common sky area. 
\textcolor{black}{Including the new confirmations discussed above, there are 117 confirmed \textit{Planck} clusters that fall within the SPT-ECS footprint, and 82 of these are associated with SPT cluster candidates at $\xi>4$.
Of the remaining 35 clusters in \textit{Planck} but not confirmed by SPT, 32 are at redshift $z < 0.25$---where the SPT filtering both reduces the completeness of the catalog and the fidelity of the mass estimates---and 5 of these 32 confirmed clusters also excluded because they are in regions excluded by the SPT point source veto.
For the 3 \textit{Planck} clusters at $z > 0.25$ but not confirmed by SPT, we find two of these systems match candidates just below the SPT-selection threshold with PSZ2 G244.74$-$28.59 (\textit{Planck} S/N=5.9, $z=0.33$) at $\xi=3.97$ and PSZ2 G251.13-78.15 (S/N=4.6, $z=0.3$) at $\xi=3.2$.
There is also radio source nearby to PSZ2 G244.74$-$28.59, which---based on the methodology of Section \ref{sec:contam}---could reduce the $\xi$ value by 0.08 to 0.7 for a source spectral index of -1 to -0.5.  }
The final unmatched cluster, PSZ2~G282.14+38.29 (S/N=4.9, $z=0.33$ with validation from Pan-STARRS) is flagged as having a nearby point source detected at 857 GHz and is measured at $\xi=-0.3$ in our sample. 
We do not detect a large excess of red-sequence galaxies in Pan-STARRS at this cluster location. 
While the SZ flux from one source (PSZ2 G244.74$-$28.59) may be diminished by the presence of a nearby radio source (which should also influence the \textit{Planck} detection) and we do not independently confirm G282.14+38.29,  we find the SPT selection to be consistent with expectations as relates to the PSZ2 sample with 39/42 of the reported \textit{Planck} clusters at $z\ge0.25$  and not in a point-source vetoed region also in the \surveyshort \ sample. Further exploration of the differences between the estimated masses for the \textit{Planck} and SPT samples  will require detailed modeling of the selection functions of the two surveys in their jointly accessible mass and redshift ranges and is beyond the scope of this work.

\begin{deluxetable}{lccc}[t]  
\tablecolumns{3}
\tablecaption{\label{tab:confirmed_planck} New confirmations of \textit{Planck} clusters.}
\tablehead{ 
\colhead{PSZ2 Name} &
\colhead{$z$} & 
\colhead{Separation ( \arcmin)}}
\startdata
PSZ2 G011.36-72.93 & 0.63$\pm$0.04 & 2.4\\
PSZ2 G011.92-63.53$^{a}$ & 0.24$\pm$0.02 & 1.1\\
PSZ2 G025.07-78.64 & 0.225$\pm$0.033 & 0.3\\
PSZ2 G029.55-60.16 & 0.218 & 2.8\\
PSZ2 G210.02-56.38 & 0.236$\pm$0.004 & 0.7\\
PSZ2 G216.76-41.84$^{b}$ & 0.39$\pm$0.01 & 1.1\\
PSZ2 G221.06-44.05 & 0.396  & 0.8\\
PSZ2 G227.61-84.72 & 0.432$\pm$0.009 & 0.7\\
PSZ2 G231.74-70.59 & 0.275$\pm$0.005 & 1.7\\
PSZ2 G240.71-74.03 & 0.40$\pm$0.01 & 0.8\\
PSZ2 G271.53+36.41 & 0.51$\pm$0.04 & 0.8\\
PSZ2 G282.11+38.61 & 0.30$\pm$0.02 & 1.6\\
PSZ2 G295.27+32.25 & 0.71$\pm$0.04 & 1.0\\
\enddata 
\tablecomments{Redshifts and angular separations (in arcminutes) from SPT cluster positions for PSZ2 \citet{planck15-27} candidates reported without redshifts that are associated with \surveyshort \ clusters. We find good agreement with the redshift reported for PSZ2~G011.92$-$63.53 by  \citet{maturi19} but find $\delta_z > 0.2$ for PSZ2~G011.36$-$72.93. We also note that we associate PSZ2~G029.55$-$60.16 with ACO S 1048 \citep{abell89}.}
\tablenotetext{a}{Associated by \citet{planck15-27} with ACO 3296, but no redshift provided}
\tablenotetext{b}{Associated by \citet{planck15-27} with ACO S 443, but no redshift provided}
\end{deluxetable}

\subsection{The \surveyshort \ Strong Lensing Subsample}

The strong gravitational lensing regime, often identified via the presence of highly magnified and multiply imaged background galaxies lensed by foreground gravitational potentials, provides a unique probe of the cores of massive structures.
Galaxy clusters have long been recognized as areas in which to productively search for strong gravitational lenses (see review by \citealt{meneghetti13} and more recent works by  \citealt{kneib11,bayliss11,lotz17,diehl17, sharon19} amongst many others). 
We examine the \surveyshort \ sample for signatures of strong lensing in the Magellan/PISCO and DES imaging data as well as in archival and dedicated observations from the  \textit{Hubble Space Telescope}, the latter from a snapshot program (PID 15307, PI: Gladders) designed to characterize the central regions of massive clusters from SPT-SZ and \surveyshort.  

\begin{figure*}[t]
\begin{center}
\includegraphics[width=7in]{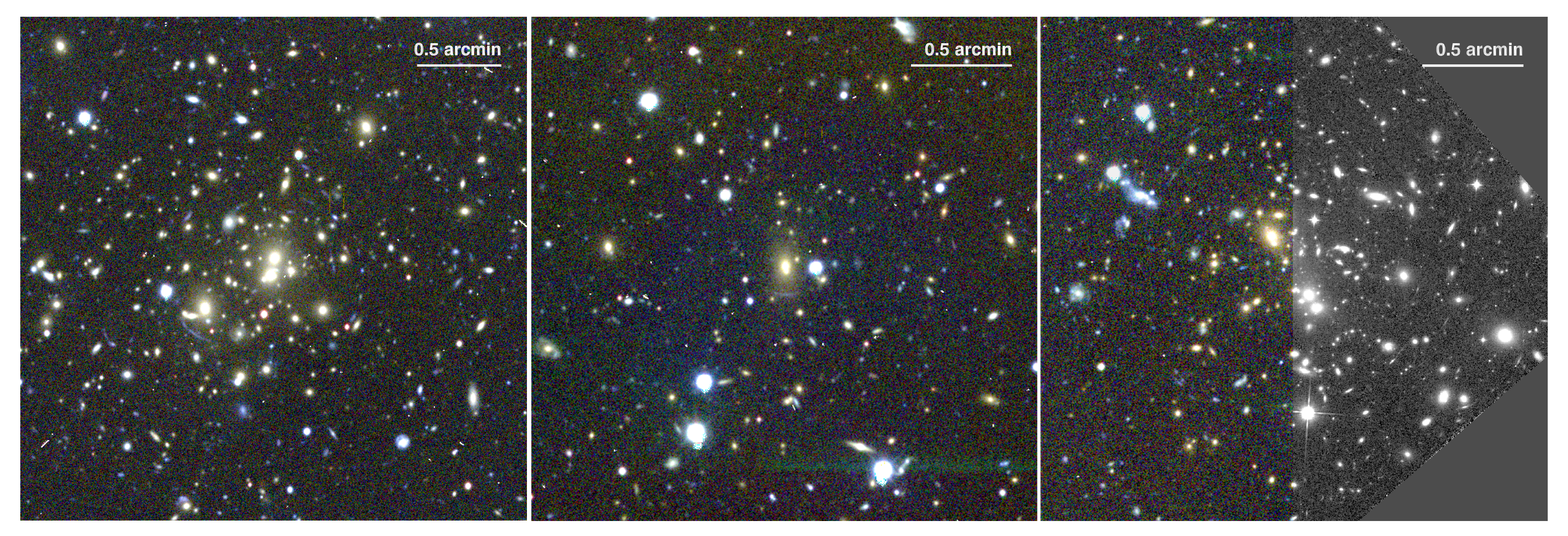}
\caption{Three strong lensing clusters from the \surveyshort \ survey. (Left) SPT-CL~J0512$-$3848 at $z=0.33$, (Middle) SPT-CL~J1223$-$3014 at $z=0.48$, (Right) SPT-CL~J0049$-$2440 at $z=0.53$ (first reported as Vidal 14; \citealt{depropris99}). In each panel we show  $\sim300$~s exposure imaging from PISCO in the \textit{gri-}bands. This data was taken in good ($<0.7\arcsec$) seeing that enables the strong lensing identification. In the right panel we also show F110W data from our ongoing \textit{HST} snapshot program. }
\label{fig:strong_lensing}
\end{center}
\end{figure*}

We find that \nstrong \  of the \surveyshort \ systems exhibit unambiguous signs of strong lensing; we flag all of these systems in Tables \ref{tab:sampletable} and \ref{tab:sample4sigma}. 
Some of these systems have been previously identified as strong lenses---see \citet{smail91,sand05,covone06,zitrin11,hamilton12,gruen14,ebeling17,ebeling18,newman18,repp18,jacobs19,petrillo19,coe19}---and in the online data for Tables \ref{tab:sampletable} and \ref{tab:sample4sigma} we also link individual previously known strong lenses to these works. 
In total over 110 systems from SPT-SZ and \surveyshort \ have been identified as strong gravitational lenses;  
a robust statistical characterization of the PISCO  and \textit{HST} data will be the subject of future work. 
In Figure \ref{fig:strong_lensing} we display high-quality PISCO data for three of the \surveyshort \ strong lenses as well as data from our \textit{HST} program for the third.

\section{The SZ properties of the joint SPT-redMaPPer cluster sample}\label{sec:sptrm}

Having constructed the \surveyshort \ cluster sample we now leverage the overlap between the DES and SPT surveys to  jointly characterize the SZ and richness properties of massive clusters in the Year 3 DES redMaPPer optically selected catalog (see Section \ref{sec:des}). 
We focus on two properties here: the richness-mass relation of these systems (a key ingredient in cosmological analyses of optical clusters that has been previously probed in numerous works e.g., \citealt{farahi16,simet17,geach17b,murata18,mcclintock19,raghunathan19a}) 
and the offsets between the SZ-based cluster centers and the optical centers as defined by the most probable central galaxy as determined by the RM algorithm. 
This distribution is useful for both cosmological studies (e.g., as an important input in weak-lensing mass calibration of clusters, \citealt{johnston07,george12,dietrich19}) and astrophysical studies, as it probes the dynamical states of clusters \citep{sanderson09,mann12,rossetti16}.  
It can also serve as a test of cluster-centering algorithms.

Following a similar criterion to \citet{saro15} we cross match the optically selected RM sample with  SZ clusters by:
\begin{itemize}
\item Rank-ordering each cluster list: for the SPT clusters by decreasing $\xi$, and for the RM clusters by decreasing $\lambda$
\item Matching each SZ system to the richest RM cluster within $\delta_z=0.1$ and projected separation between the SZ and RM center $<1.5$ Mpc at the cluster redshift and then 
\item Removing each matched RM cluster from the possible matching pool and continuing the process until the last SZ cluster has been checked for a match. 
\end{itemize}
Note that we do \textit{not} compute a probability of random association here for each SZ cluster in this list as we have already statistically identified a high-probability association between a cluster detected by the RM algorithm run in ``scanning'' mode ( Section \ref{subsec:desmatch})  and the SZ detection. The matching criterion we've chosen in this selection allows us to more fully capture the properties of the RM algorithm when it is run in its standard, blind-search mode; in particular clusters that scatter low in richness in the blind search are not cut from this analysis.
This procedure is also repeated for the full SPT-SZ sample (updating the \citealt{saro15} results which centered on the DES Science Verification Region). 
We confirm 13 new clusters at $\xi>4.5$ via this method, the majority of which are above the redshift limits reported in B15 (though we found some of these limits were  overestimated in cases of poor seeing). 
The new clusters are reported in Table \ref{tab:newsptsz} and we note that the sample of $\xi<4.5$ SPT-SZ systems will be discussed in detail in M. Klein et al. (in preparation).

\begin{figure}[t]
\hspace*{0.0in}
\vspace*{-0.0in}
\begin{center}
\includegraphics[width=3.3in]{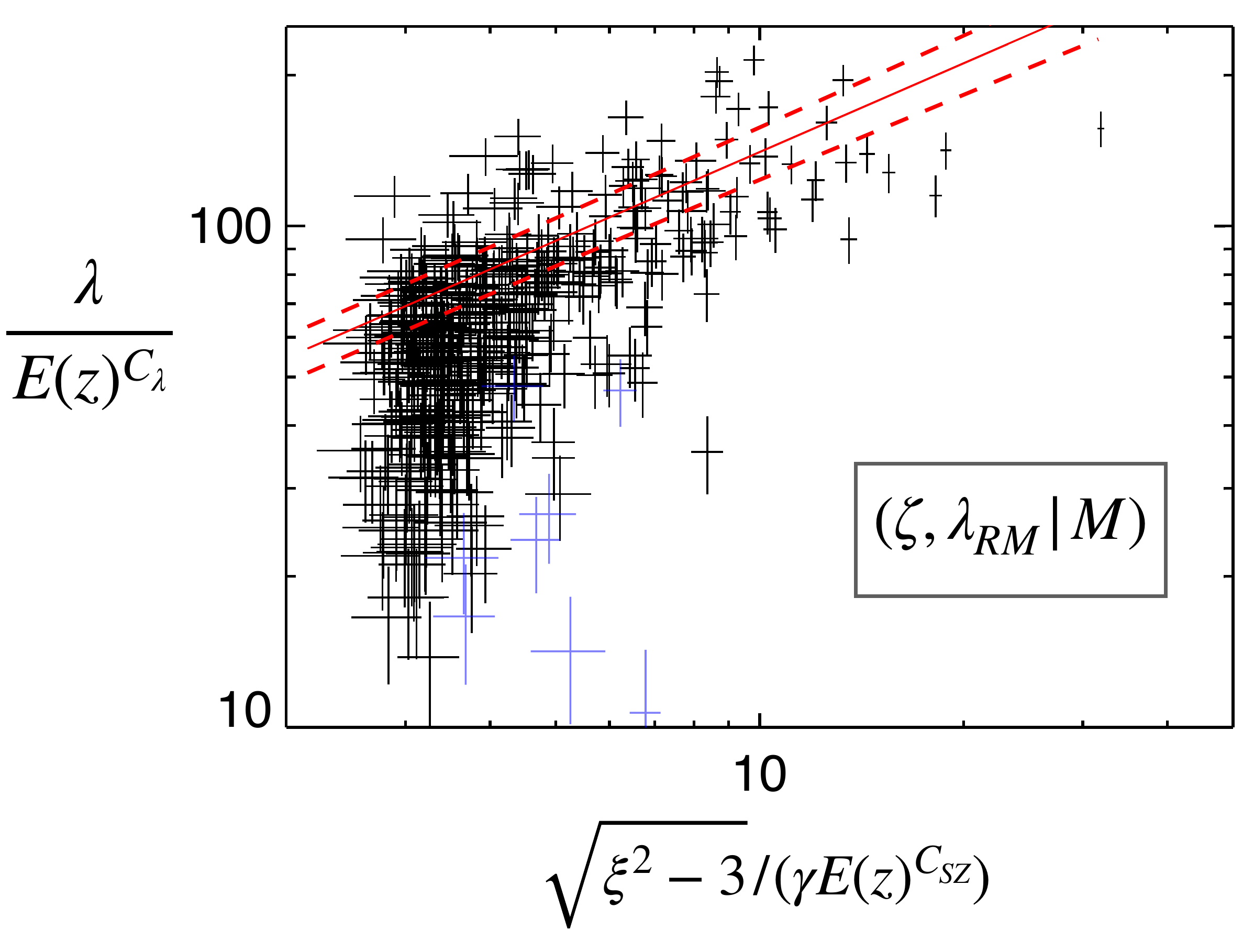}
\caption{Richness versus normalized $\xi$ values for the ensemble SPT-RM volume-limited cluster sample; light blue points are clusters for which $>30\%$ of the DES data was masked in the vicinity of the SPT cluster. The $\xi$ values are normalized by the field scaling factors discussed in Section \ref{subsec:szmass}. Overplotted in red is the best-fit $\lambda-\xi$ relation as calculated in Section \ref{subsec:lambdam}.}
\label{fig:lambda_sz}
\end{center}
\end{figure}

Including SZ cluster candidates detected at $\xi>4.5$ in the SPT-SZ survey \citep{bleem15b} we find \nredmapperspt \ clusters in the ensemble SPT-RM cluster sample.
Limiting the redshift range to $z>0.25$
reduces the sample to \nredmapperzcut \ systems, and to the volume-limited catalog  results in a sample of 249 (\nredmapperzcutvol) clusters at $\xi\ge5(4)$; the richness versus $\xi$ (normalized for the field scaling factors, see \ref{subsec:szmass}) are shown in Figure \ref{fig:lambda_sz}.  

\subsection{The Richness--Mass Relation of SPT-RM Clusters}\label{subsec:lambdam}

We use the optical richness ($\lambda$) measurements of SPT clusters matched to the Y3 RM catalog to calibrate the richness--mass relation, taking the SPT selection into account. Assuming our fiducial fixed cosmology, we simultaneously constrain the SZ scaling relation parameters through the number counts of the SPT cluster sample (as discussed in Section~\ref{sec:abundance}) and the parameters of the richness scaling relation.
This analysis follows \citet{saro15} with the exception that we now also account for the effects of correlated scatter among $\zeta$ and richness.

\subsubsection{Richness--Mass Relation: Likelihood Function}

Along with the $\zeta$--mass relation defined above in Eq.~\ref{eqn:zetam}, we define the richness--mass relation
\begin{equation} \label{eq:lambda_mass}
\begin{split}
   \langle\ln\lambda\rangle =& \ln A_\lambda + B_\lambda \ln\left(\frac{M_{500c}}{3\times10^{14}M_\odot h^{-1}}\right) \\
   &+ C_\lambda\ln\left(\frac{E(z)}{E(z=0.6)}\right).
\end{split}
\end{equation}
A covariance matrix describes the correlated intrinsic scatter between the two observables $\zeta$ and $\lambda$
\begin{equation}
    \Sigma_{\zeta-\lambda} = \begin{pmatrix} \sigma_{\ln\zeta}^2 & \rho_{\mathrm{SZ}-\lambda}\sigma_{\ln\zeta}\sigma_{\ln\lambda} \\
    \rho_{\mathrm{SZ}-\lambda}\sigma_{\ln\zeta}\sigma_{\ln\lambda} & \sigma_{\ln\lambda}^2 + \lambda^{-1}
    \end{pmatrix}.
\end{equation}
The contribution $\lambda^{-1}$ to the intrinsic scatter in richness represents the Poisson noise in the number of member galaxies observed at a fixed cluster mass.
We note that we expect positive correlation in the scatter between $\zeta$ and $\lambda$ as both are projected quantities (see e.g., \citealt{angulo12}). 

The joint scaling relation then reads
\begin{equation}
    P\Bigl(\begin{bmatrix}\ln\zeta \\ \ln\lambda\end{bmatrix}|M,z\Bigr) = \mathcal N\Bigl(\begin{bmatrix}\langle\ln\zeta\rangle(M,z) \\ \langle\ln\lambda\rangle(M,z)\end{bmatrix}, \Sigma_{\zeta-\lambda}\Bigr).
\end{equation}

Following \citet{bocquet19}, the likelihood function for our number counts and richness calibration analysis is
\begin{equation}\begin{split}
\label{eq:likelihood_SPT_RM}
\ln \mathcal L(\vec p) =  & \sum_i \ln \frac{dN(\xi_i, z_i| \vec p)}{d\xi dz} \\
&- \int_{z_\mathrm{cut}}^\infty dz \int_{\xi_\mathrm{cut}}^\infty d\xi \frac{dN(\xi, z| \vec p)}{d\xi dz} \\
&+ \sum_j \ln P_j(\mathrm{match}) P(\lambda_{\mathrm{obs},j}^{>5} | \xi_j, z_j, \vec p)
\end{split}\end{equation}
up to a constant, where the first sum runs over all clusters $i$ in the SPT sample above $\xi>5$ and $z>0.25$ and the second sum runs over all SPT clusters $j$ for which a RM richness measurement is available.
Note that the first two lines represent the number-count likelihood defined earlier in Equation~\ref{eq:like_NC}.
The term $P(\mathrm{match})=1-P(\mathrm{random})$ describes the excess probability of matching a RM cluster to an SPT cluster over random associations $P(\mathrm{random})$. 
The other term in the last line is computed as
\begin{equation}
\begin{split}
 P(\lambda_\mathrm{obs} | \xi, z, \vec p) =& \iiint dM\, d\zeta\, d\lambda\, \left [ \right. \\
 &P(\lambda_\mathrm{obs}|\lambda)
 P(\xi|\zeta) \\
 & P(\zeta,\lambda|M,z,\vec p) P(M|z,\vec{p})
   \left. \right ].
\end{split}
\end{equation}
Finally, we account for the richness cut $\lambda_\mathrm{obs}>5$ in the volume-limited redMaPPer catalog and evaluate
\begin{equation}
P(\lambda_\mathrm{obs}^{>5} | \xi, z, \vec p) = \frac{\Theta(\lambda_\mathrm{obs} >5) P(\lambda_\mathrm{obs} | \xi, z, \vec p)}
{\int_5^\infty d\lambda_\mathrm{obs} \Theta(\lambda_\mathrm{obs} >5) P(\lambda_\mathrm{obs} | \xi, z, \vec p)}
\end{equation}
with the step function $\Theta$.

\subsubsection{Richness--Mass Relation: Results}
With this machinery in place we are now ready to explore the mass-richness relation of the SPT-RM sample.
Assuming our fiducial cosmology, we evaluate the likelihood presented in Eq.~\ref{eq:likelihood_SPT_RM} of the SPT cluster number counts (which constrains the SZ scaling relation parameters), and the likelihood of the RM richnesses (which constrains the RM richness scaling relation parameters).
We only use the SPT-SZ sample for the SPT number counts (to enable an independent constraint on $\gamma_\textrm{ECS}$, described below) but we use redMaPPer richnesses for the full SPT-SZ+\surveyshort \  sample. 
\begin{figure}
\includegraphics[width=\columnwidth]{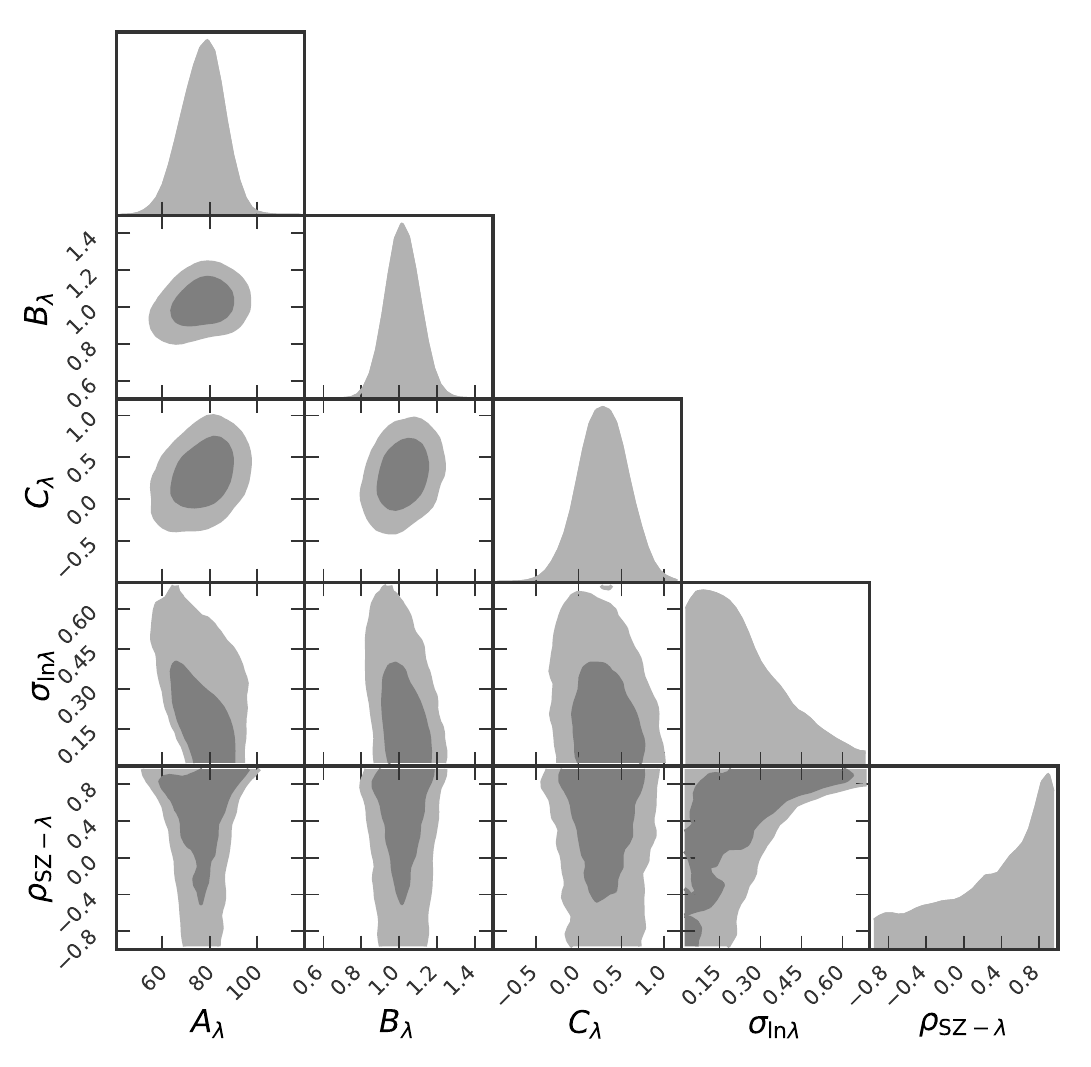}
\caption{Parameters of the richness--mass relation defined in Eq.~\ref{eq:lambda_mass} and the correlation coefficient,  $\rho_{\mathrm{SZ}-\lambda}$,  between the SZ signal ($\zeta$) and richness.}
\label{fig:scal_rel_params}
\end{figure}

\begin{deluxetable}{cc}
\tablecaption{\label{tab:tablelm}Parameters of the Richness-Mass Relation.}
\tablehead{\colhead{\hspace{0.75cm}Parameter \hspace{0.75cm}} & \colhead{\hspace{1.5cm}Constraint}\hspace{1.5cm}}
\startdata
$A_\lambda$ & \textcolor{black}{$76.9\pm8.2$} \\
$B_\lambda$ & \textcolor{black}{$1.020\pm0.080$} \\
$C_\lambda$ & \textcolor{black}{$0.29\pm0.27$} \\
$\sigma_{\ln\lambda}$ & \textcolor{black}{$0.23\pm0.16$} \\
$\rho$ & \textcolor{black}{$>-0.78$ (95\% CL)}
\enddata
\tablecomments{The richness--mass relation is defined in Eq~\ref{eq:lambda_mass}. We also quote the constraint on the correlation coefficient between the scatter in the SZ signal and richness $\rho_{\mathrm{SZ}-\lambda}$. The constraints are obtained using redMaPPer matches to the $\xi>4.5, z>0.25$ SPT sample.}  
\end{deluxetable}

We present the constraints on the richness--mass relation in Figure~\ref{fig:scal_rel_params} and in Table~\ref{tab:tablelm}. 
Compared to previous constraints using 19  clusters from SPT-SZ at $\xi>4.5$ in the DES Science Verification region \citep{saro15}, we find a normalization that is $\sim1.2\sigma$ higher, with the slope and redshift evolution consistent.

We also  compare against the DES weak lensing analysis of the Year 1 RM sample reported in \citet{mcclintock19}, which was also analyzed at our fiducial cosmology. Note that the DES weak lensing analysis constrains $P(M_{200\mathrm{m}}|\lambda)$---with masses defined with respect to the \textit{mean} density of the Universe---whereas our analysis constrains $P(\lambda|M_{500\mathrm{c}})$. We convert $M_{500\mathrm{c}}$ to $M_{200\mathrm{m}}$ assuming a Navarro, Frenk and White (NFW; \citealt{navarro96}) profile and the concentration--mass relation from \citet{child18}.\footnote{We use the Colossus package \url{https://bitbucket.org/bdiemer/colossus}} We invert our relation as
\begin{equation}
P(M_{200\mathrm{m}}|\lambda) = \int dM_{200\mathrm{m}}P(\lambda|M_{200\mathrm{m}})P(M_{200\mathrm{m}})
\end{equation}
with the halo mass function prior $P(M_{200\mathrm{m}})$.

\begin{figure}
\includegraphics[width=\columnwidth]{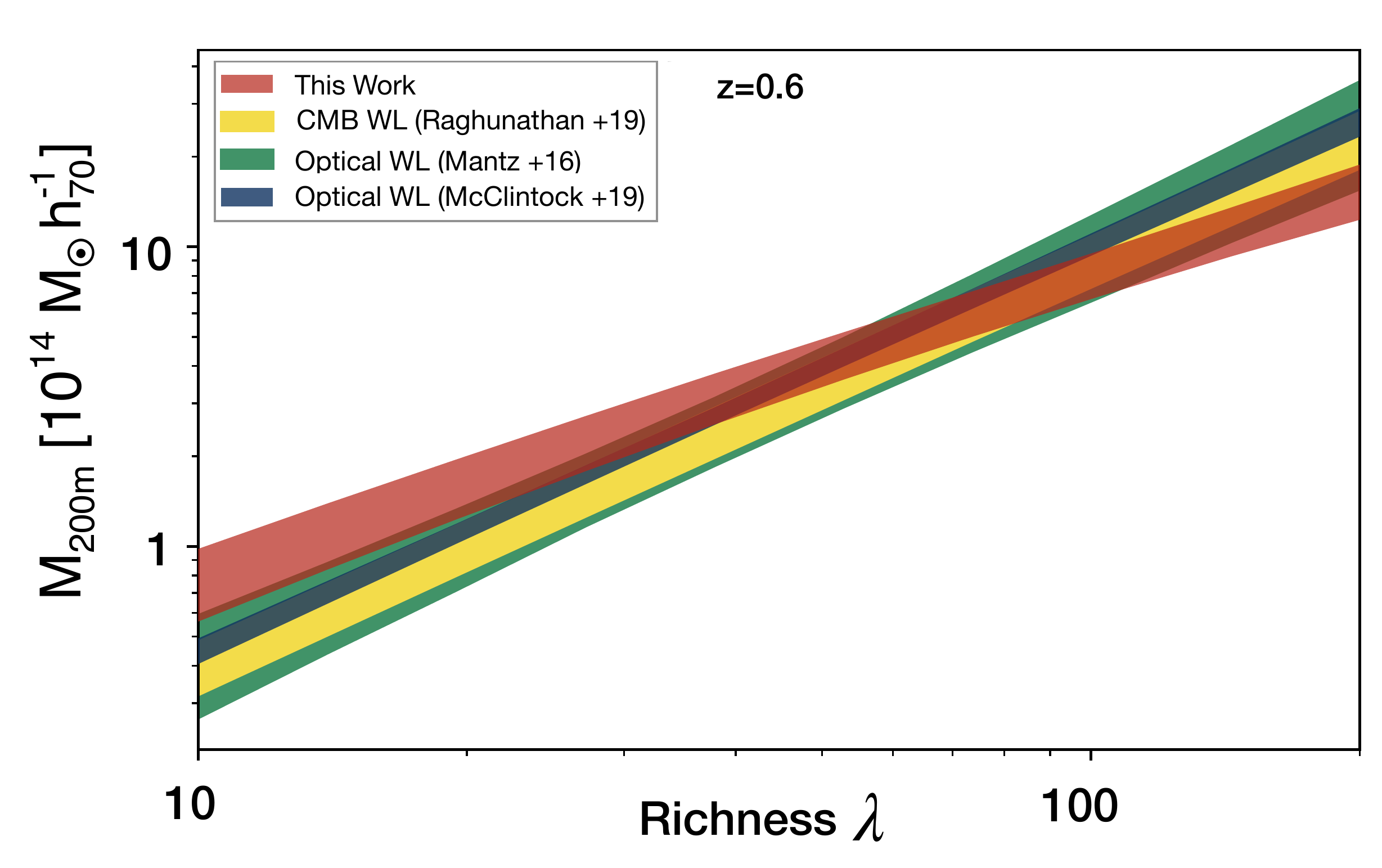}
\caption{The mass--$\lambda$ relation evaluated at our pivot redshift $z=0.6$ determined from SPT cluster number counts assuming our fiducial  $\Lambda$CDM cosmology.
We convert our $\lambda-M_{500\mathrm{c}}$ relation to $M_{200\mathrm{m}}-\lambda$ for ease of comparison with the literature. The relations calibrated from DES Y1 shear or CMB lensing (the latter driven by an informative prior) favor steeper slopes, but there is good agreement at their pivot richness $\lambda_0=40$ \citep{mcclintock19, raghunathan19a}.
The calibration from an X-ray selected sample with optical weak-lensing provides a richness--mass relation that is very similar to the DES Y1 shear result \citep{mantz16}.}
\label{fig:lambda_mass}
\end{figure}

In Figure~\ref{fig:lambda_mass}, we show the mass--lambda relation from our work and examples from the literature.  At our scaling relation pivot redshift ($z=0.6$, see Equation \ref{eq:lambda_mass}), the scaling relation normalizations are consistent at $\lambda\approx60$ or $M_{200\mathrm{m}}\approx5\times10^{14}M_\odot$. However, there are some visible differences in the slope.
We approximate\footnote{Strictly speaking, we compare the slopes of the $\lambda$--mass and the mass--$\lambda$ relations. We checked that the conversion of our relation to mass--$\lambda$ mostly shifts the amplitude of the relation while leaving the slope almost unchanged.} the slope $F_\lambda$ in our $P(M_{200\mathrm{m}}|\lambda)$ relation as
\begin{equation}
F_\lambda \equiv 1/B_\lambda = 0.981 \pm 0.077.
\end{equation}

We find our slope is $\sim$30\% shallower than the slope from the DES Y1 analysis \citep[$F_\lambda =1.356\pm0.052$;][]{mcclintock19}, with a 4$\sigma$ offset between the two constraints.
To reproduce the slope of the \cite{mcclintock19} relation we would require a significant shift in our assumed cosmology along the $\Omega_\mathrm{m}$ and $\sigma_8$ degeneracy direction (see e.g., \citealt{costanzi18}); however a full cosmological interpretation is beyond the scope of this work, and would 
depend on fully accounting for selection effects in the RM sample under study as well as on degeneracies and covariances in a wider multi-dimensional parameter space.

A weak-lensing analysis using data from the Sloan Digital Sky Survey (SDSS) finds an amplitude and slope that are consistent with \cite{mcclintock19} at better than $1\sigma$ \citep{simet17}.
Another weak-lensing study using SDSS data finds a much shallower slope centered at $\lambda\propto M^{0.64}$ using lensing alone; this slope becomes consistent with unity---and thus our measurement---when combining lensing and cluster abundance \citep{murata18}.
Qualitatively similar results are presented in an analysis of the richness--mass relation using first-year HSC data \citep{murata19}.
A weak-lensing calibration of an X-ray selected cluster sample yields constraints on the richness--mass relation that are centered on the results from \cite{mcclintock19}, but with large uncertainties \citep{mantz16}.

Moving beyond optical weak lensing, two calibrations of the mass--$\lambda$ relation using lensing of the CMB \citep{baxter18, raghunathan19a} recover amplitudes of the mass--$\lambda$ relation that are compatible with both the calibration from DES Y1 shear measurements  and this work. Note however, that the slope parameters were not constrained by the CMB lensing measurements, where informative priors were applied.
The richness-mass relation has also been calibrated using the clustering of clusters \citep{baxter16} and the measurement of pairwise velocity dispersions \citep{farahi16}; both methods show consistency at the $2\sigma$ level. Finally, a study of the phase-space of galaxy dynamics provides a calibration of the richness--mass relation with a slope that is consistent with unity at $<1\sigma$ \citep{capasso19}. However, their relation exhibits  strong redshift evolution, which leads to an offset in the relations at our pivot redshift $z=0.6$.

\subsubsection{Richness--Mass Relation: Constraint on $\gamma_\textrm{ECS}$}

By using only the SPT-SZ data in the number counts we can also use this test to independently evaluate our estimated value for $\gamma_\textrm{ECS}$ presented in Section \ref{sec:abundance}.
 We obtain a calibration of the correction factor $\gamma_\mathrm{ECS}$ between SPT-SZ and \surveyshort \  
\begin{equation}
\gamma_\mathrm{ECS} = 1.054 \pm 0.075.
\end{equation}
This determination of $\gamma_\mathrm{ECS}$ is different from the result presented above in section~\ref{sec:abundance}. In both cases, $A_\mathrm{SZ}$ is constrained to yield number counts from SPT-SZ that match our fixed fiducial cosmology. In section~\ref{sec:abundance}, $\gamma_\mathrm{ECS}$ was calibrated by also demanding that the \surveyshort \  number counts match that cosmology---any relative offset in the amplitude of the $\zeta$--mass relation between SPT-SZ and \surveyshort \  is thus absorbed by $\gamma_\mathrm{ECS}$. In the calibration presented here, the redMaPPer richnesses serve as the relative anchor between the SPT-SZ and \surveyshort \  surveys. 
The two determinations of $\gamma_\mathrm{ECS}$ agree at the $0.8\sigma$ level. We conclude that our empirical modeling of the full SPT-SZ+\surveyshort \  sample with an overall amplitude offset is adequate and when reporting cluster masses we adopt the  mean recovered constraint from the more precise NC analysis result as our default.

\subsection{redMaPPer-SZ center offset distribution}

We next explore the distribution of separations between the redMaPPer and SPT-determined cluster centers. 
Based on visual inspection of $>100$ matched X-ray and RM clusters in SDSS, \citet{rozo14b} found that the gas centers and central galaxies should be well aligned (within 50 kpc) 80\% of the time, with it being rare to find a separation of $>300$ kpc between the two (results consistent with previous findings by e.g., \citealt{lin04b}). 
In the SDSS sample, the RM algorithm selected the visually identified central galaxy  in $86\pm3\%$ of systems and had a long uniform tail to 800 kpc for the remainder of systems. 
These gas-central galaxy separations were further quantified for RM clusters by \citet{saro15} with 19 SPT-RM clusters in DES Science Verification data and \citet{zhang19} for 144 (67) systems in SDSS (DES); the latter analysis using archival \textit{Chandra} X-ray data as analyzed in \citet{hollowood18}. 
With differing model parameterizations these works found $\sim 63-84\%$ of all clusters to be well-centered. 

Following these previous works, we adopt two different models for this offset distribution for the SPT-RM sample, one modeling offsets relative to the cluster mass scale (via $R_{500c}$) and the other 
relative to a cluster extent that scales as a function of RM galaxy richness.
Both of these models assume that the offset distribution can be modeled as a central core of well-aligned clusters with small separation combined with a subdominant population of clusters with large offsets. 
Physically this corresponds to the cluster population being composed of a mixture of relaxed and merging clusters with some additional scatter introduced via possible misidentification of central galaxies by the RM algorithm. 

The dynamical state of the cluster population is also traced by the morphology of the cluster gas, which can be measured by X-ray observations (note the filtering applied to the SPT maps makes it difficult to extract a robust gas morphological measurement from the SZ data). 
The X-ray morphology has been measured via the $A_\textrm{phot}$ statistic for 50 of the SPT-RM clusters that are also part of a \textit{Chandra} X-ray Visionary Project  (XVP; PI: Benson, \citealt{nurgaliev17}); 38 of the systems in the \citet{zhang19} DES Y1 analysis mentioned above are part of the SPT-XVP. 
 The $A_\textrm{phot}$  statistic is a quantification of the amount of azimuthal asymmetry present in the X-ray photon count distribution and has been shown to be a robust morphological measure even when used on X-ray data with a relatively low number of counts ($\sim2000$ counts/cluster)  such as the SPT-XVP observations \citep{nurgaliev13}. 
  We plot as an inset in Figure \ref{fig:offset_model1fit} the SZ-optical-offset distribution of these 50 clusters.  
The outlier in this inset plot is SPT-CL~J2331$-$5051($A_\textrm{phot}=0.14$) which may be captured pre-merger with SPT-CL~J2332$-$5053. The RM algorithm has selected what appears to be the central galaxy of the latter cluster and found a smaller structure of $\lambda=8$ at the location of SPT-CL~J2331$-$5051 which is the more massive system inferred from both the SZ and X-ray observations  (see further discussion of this system in \citealt{andersson11} and \citealt{huang19}). 
The  $A_\textrm{phot}$  distribution is a continuum, but adopting the somewhat arbitrary choice of \citet{mcdonald17} with  $A_\textrm{phot} < 0.1$ classified as ``relaxed'' (17 systems) and $A_\textrm{phot} > 0.5$ as ``disturbed'' (10 systems) we find the median offset of the relaxed (disturbed) systems to be $0.067^{+0.005}_{-0.02}R_{500c}$  ($0.23^{+0.01}_{-0.04}R_{500c}$), with the relaxed systems having a closer alignment between the SZ center and the RM most probable central galaxy, as expected.

\subsubsection{Offset Distribution relative to $R_{500c}$}

\begin{figure*}[t]
\begin{center}
\includegraphics[width=7in]{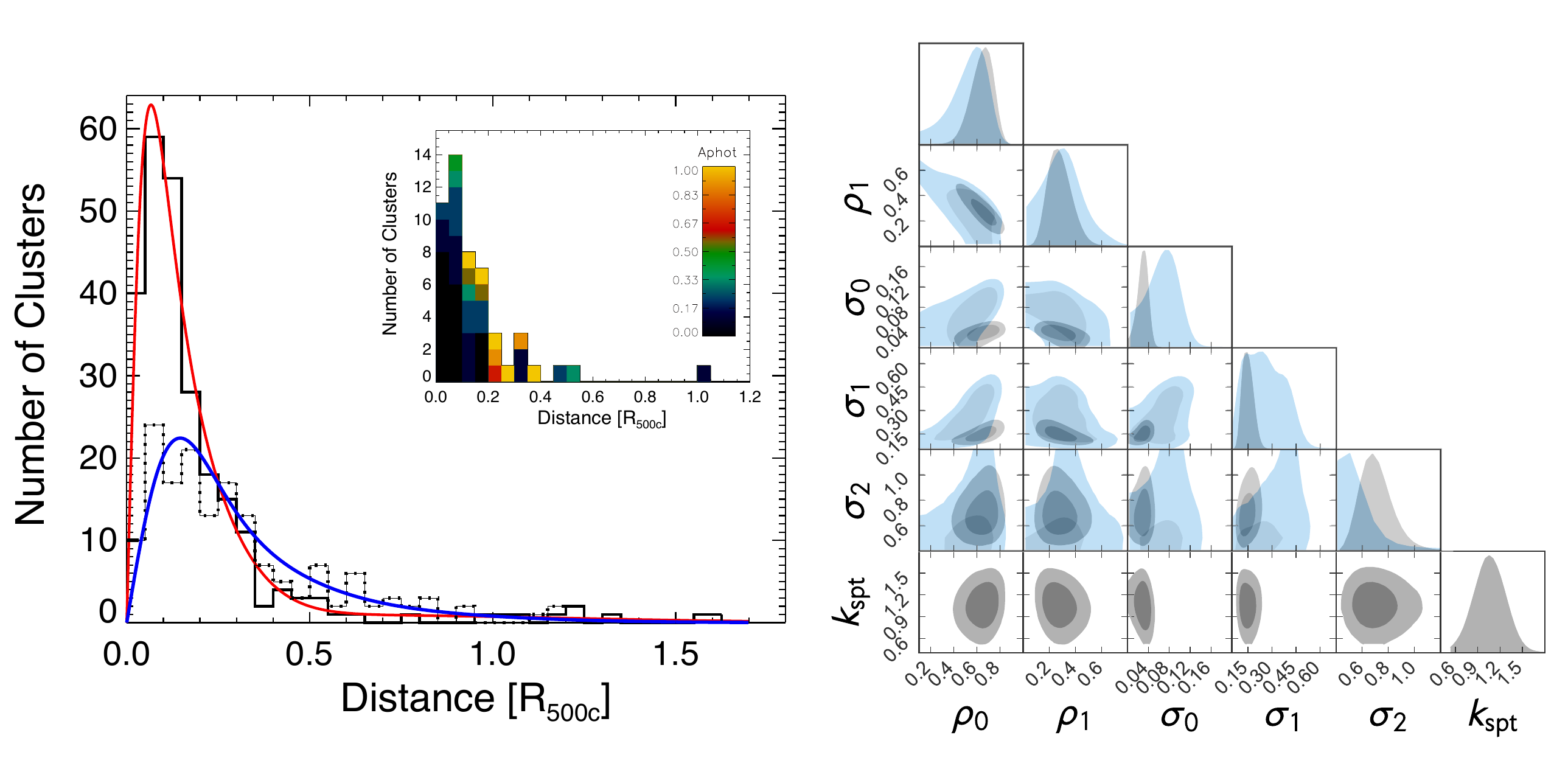}
\caption{(Left) The distribution of offsets between SPT centroids and RM most probable centers as a fraction of R$_{500c}$ for SPT systems in the RM volume-limited sample. The sample at $\xi\ge5$ is plotted in solid black and at $4<\xi<5$ with a dotted line. Overplotted in red (blue) is the best fit to the model given by Equation \ref{eq:pofx} for the high (low) significance sample. Inset is the offset-distribution of the 50 SPT-RM clusters for which the X-ray morphology statistic $A_\textrm{phot}$ has been measured. As expected the more relaxed systems (with smaller $A_\textrm{phot}$ values) on average have less spatial separation between the central galaxy and gas center. (Right) Constraints on the parameters of the offset probability distribution model. Best-fit values are given in Table \ref{tab:offset_model1tab}.  As the lower-significance sample does not have the power to constrain $\kappa_\textrm{SPT}$ we fix its value to the best-fit value from the higher-significance sample for this analysis. }
\label{fig:offset_model1fit}
\end{center}
\end{figure*}

We first consider the offset distribution relative to the cluster scale $R_{500c}$.  
For this analysis  we split the cluster population into two parts: a high-significance subset with $\xi \ge 5$ (249 clusters, median $\lambda=81$) which is the threshold  used for SPT cosmological analyses (see e.g., \citealt{bocquet19}), 
and a lower-significance sample at $4<\xi<5$ (161 systems, median $\lambda=55$). 
In Figure \ref{fig:offset_model1fit} we plot the distribution of separations between the SZ centroids and RM central galaxies for these two subsamples.

\begin{figure}[t] 
\includegraphics[width=3.5in]{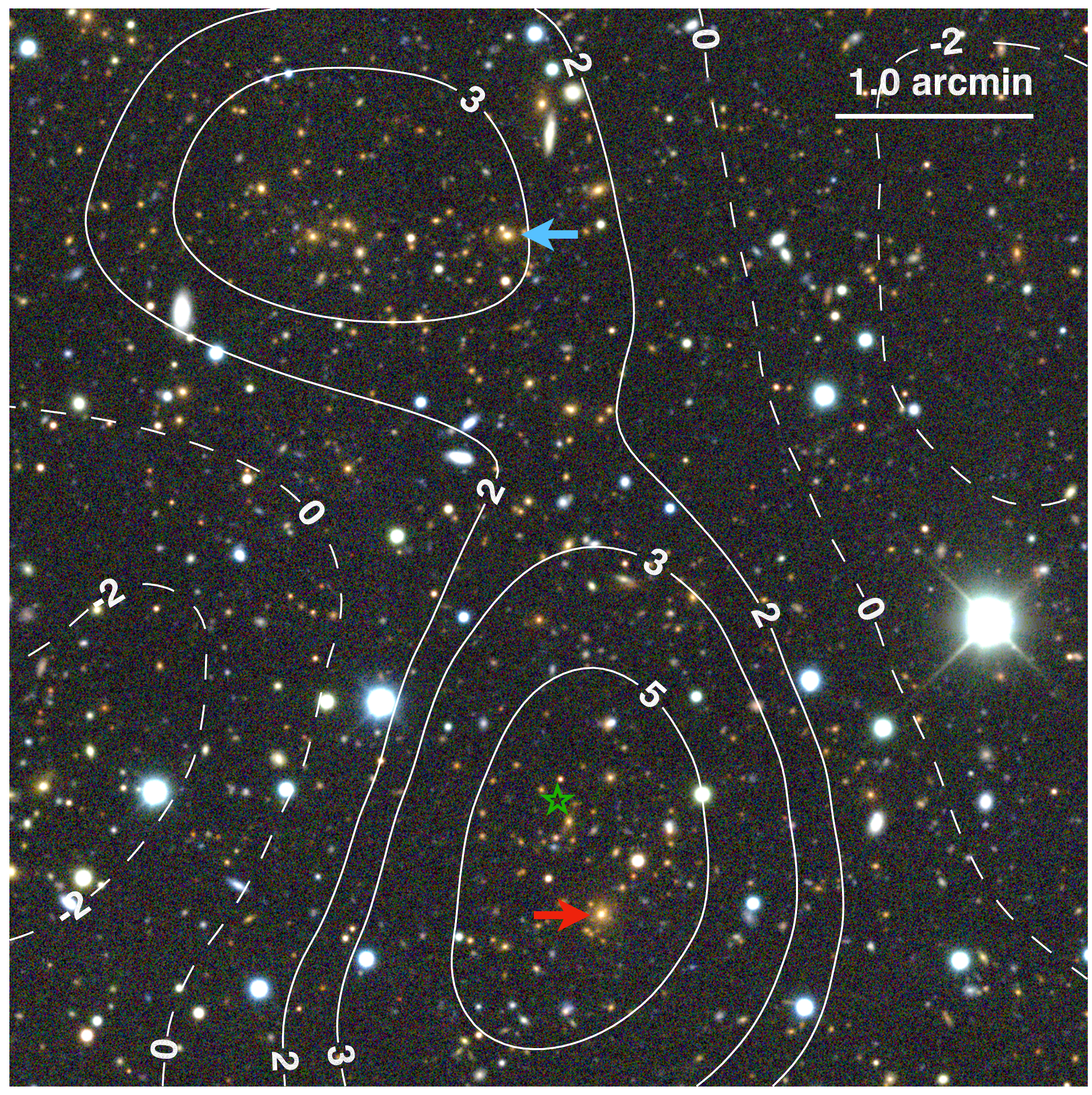}
\caption{SPT-CL~J0543$-$4250, an illustration of the small fraction of systems where the separation between RM most probable center (identified by blue arrow, $\lambda=136, z=0.609\pm0.008$) and the SZ center is greater than the RM cluster scale $R_\lambda$. The RM algorithm does identify a smaller group at the same redshift (red arrow, $\lambda =14$), significantly closer to the SPT location (green star).  Shown is DES \textit{g-,r-,i-}band imaging overlaid with SPT-SZ matched filter detection contours.}
\label{fig:merging_cluster}
\end{figure}

To characterize this distribution our model follows that of \citet{saro15}.
We have added a third Gaussian term to account for the long  tail to large separations. 
As noted above such a tail was also previously seen in analyses of SDSS clusters (and given its small sample size, the absence of a significant tail in  \citealt{saro15} is unsurprising). 
Examination of clusters with the largest separations revealed systems where the RM algorithm identified a bright galaxy near what was the lesser of two SZ peaks in merging clusters (see e.g., Figure \ref{fig:merging_cluster} and the discussion of SPT-CL~J2331$-$5051 above),  rich systems split into multiple detections (i.e., ``mispercolation'', see discussion in \citealt{hollowood18}), systems with significant masking of the optical data near the SPT position, and---for a few of the lower-significance clusters---systems with higher (but still less than 5\%) chance of random association between the SZ candidate and RM cluster.

We write the probability distribution as a function of the fractional separation, $x=r_{\textrm{offset}}/R_{500c}$ as: 
\begin{equation}\label{eq:pofx}
\begin{split}
P(x) = 2\pi x \Big(  \frac{\rho_0}{2\pi \sigma_{0}^2}e^{\frac{-x^2}{2\sigma_{0}^2}} +  \frac{\rho_1}{2\pi \sigma_{1}^2}e^{\frac{-x^2}{2\sigma_{1}^2}}  + 
\\\frac{(1-\rho_0-\rho_1)}{2\pi \sigma_{2}^2}e^{\frac{-x^2}{2\sigma_{2}^2}} \Big) 
\end{split}
\end{equation}
convolved with the SPT positional uncertainty. 
The SPT positional uncertainty is given by the cluster detection significance, $\xi$, and detection scale, $\theta_c$ 
\begin{equation}\label{eq:sptpos}
\sigma_{SPT} = \frac{\sqrt{\theta_\textrm{beam}^2 + (\kappa_\textrm{SPT}\theta_{c})^2}}{\xi}
\end{equation} 
convolved with a general astrometric uncertainty of 4-6\arcsec \ (see Section \ref{sec:processing} and W. Everett et al., (2019, in preparation)). 
where $\theta_\textrm{beam} =1.3\arcmin$ is a combination of the 95+150 GHz beams and $\kappa_\textrm{SPT}$ is a parameter of order unity \citep{story11,song12b}.

We use the {\tt emcee} package in Python \citep{foreman13} to conduct a Markov Chain Monte Carlo (MCMC) maximum likelihood analysis adopting priors of 
\begin{subequations}
\begin{align*}
		0 \leq  & \rho \leq  1.0 \\
		0 \leq & \rho_0-\rho_1 \leq 1 \\
		0 \leq & \sigma_0 \leq 0.3 \\
		\sigma_0 < & \sigma_1<  2 \\
		\sigma_1 <  & \sigma_2 < 3  \\
		0.5 \leq & \kappa_\textrm{SPT} \leq 2.
\end{align*}
\end{subequations}
Results for this parameterization for both samples are shown in Figure \ref{fig:offset_model1fit} and reported in Table \ref{tab:offset_model1tab}.
We note that the lower-significance sample does not have the power to constrain $\kappa_\textrm{SPT}$ and so we fix it to the best-fit value from the $\xi \ge 5$ sample. 

For the high-significance,  $\xi \ge5$ sample we find that the fraction of clusters in the well-centered component in this version of RM is consistent with  \citet{saro15} ($\rho_0 = 67^{+6}_{-8}\%$ vs.  $63^{+15}_{-25}\%$) with the uncertainty reduced a factor of 2 in this work. The width of this component is slightly smaller ($\sigma_0 = 0.02 \pm 0.01$ vs. $0.07 \pm 0.02$) and the width of the second component is also smaller ($\sigma_1=0.15 \pm 0.03$ versus $0.25\pm 0.07$), though we note that some of this spread is absorbed in the third Gaussian term that captures the offsets to high $R_{500c}$. 

Turning to the lower significance sample, we find it overall less well-centered than the higher significance sample, but with the parameters also less well constrained.  
Future studies using SZ clusters from the 500d \sptpol \ survey or from SPT-3G will significantly increase the number of lower-mass clusters in our SZ-matched sample and will allow us to more robustly explore miscentering trends as a function of mass. 

\begin{deluxetable}{ccc}[t]  
\tablecolumns{3}
\tablecaption{\\ Miscentering Model 1 Fits \label{tab:offset_model1tab}}
\tablehead{ 
\colhead{Parameter} &
\colhead{$\xi \ge 5$} & 
\colhead{$4 < \xi < 5$} 
 } 
 \startdata 
$\hspace{0.5cm} \rho_0    \hspace{.1cm} $ & $\hspace{1cm} 0.675^{+0.07}_{-0.08} $ & $0.54^{+0.13}_{-0.20}$ \\
$\hspace{0.5cm} \sigma_0 \hspace{.1cm}  $ & $\hspace{1cm} 0.02^{+0.01}_{-0.01} $ & $ 0.065^{+0.03}_{-0.035} $\\ 
$\hspace{0.5cm} \rho_1    \hspace{.1cm} $ & $\hspace{1cm} 0.25^{+0.08}_{-0.06} $ & $ 0.29^{+0.14}_{-0.16} $\\ 
$\hspace{0.5cm} \sigma_1 \hspace{.1cm}$  & $\hspace{1cm} 0.15^{+0.03}_{-0.03} $ & $ 0.24^{+0.14}_{-0.11} $\\ 
$\hspace{0.5cm} \sigma_2 \hspace{.1cm}$  & $\hspace{1cm} 0.70^{+0.125}_{-0.09} $ & $ 0.48^{+0.15}_{-0.07} $\\ 
$\hspace{0.5cm} \kappa_\textrm{SPT}   \hspace{.1cm}  $ & $\hspace{1cm} 1.0 \pm 0.2$ & $ - $ 
\enddata 
\tablecomments{Best-fit miscentering parameters for the SPT-RM Volume Limited Sample as characterized in Equation \ref{eq:pofx}. }
\end{deluxetable}

\subsubsection{Offset Distribution relative to $R_\lambda$}\label{subsec:descentering}
\begin{figure*} 
\begin{center}
\includegraphics[width=7in]{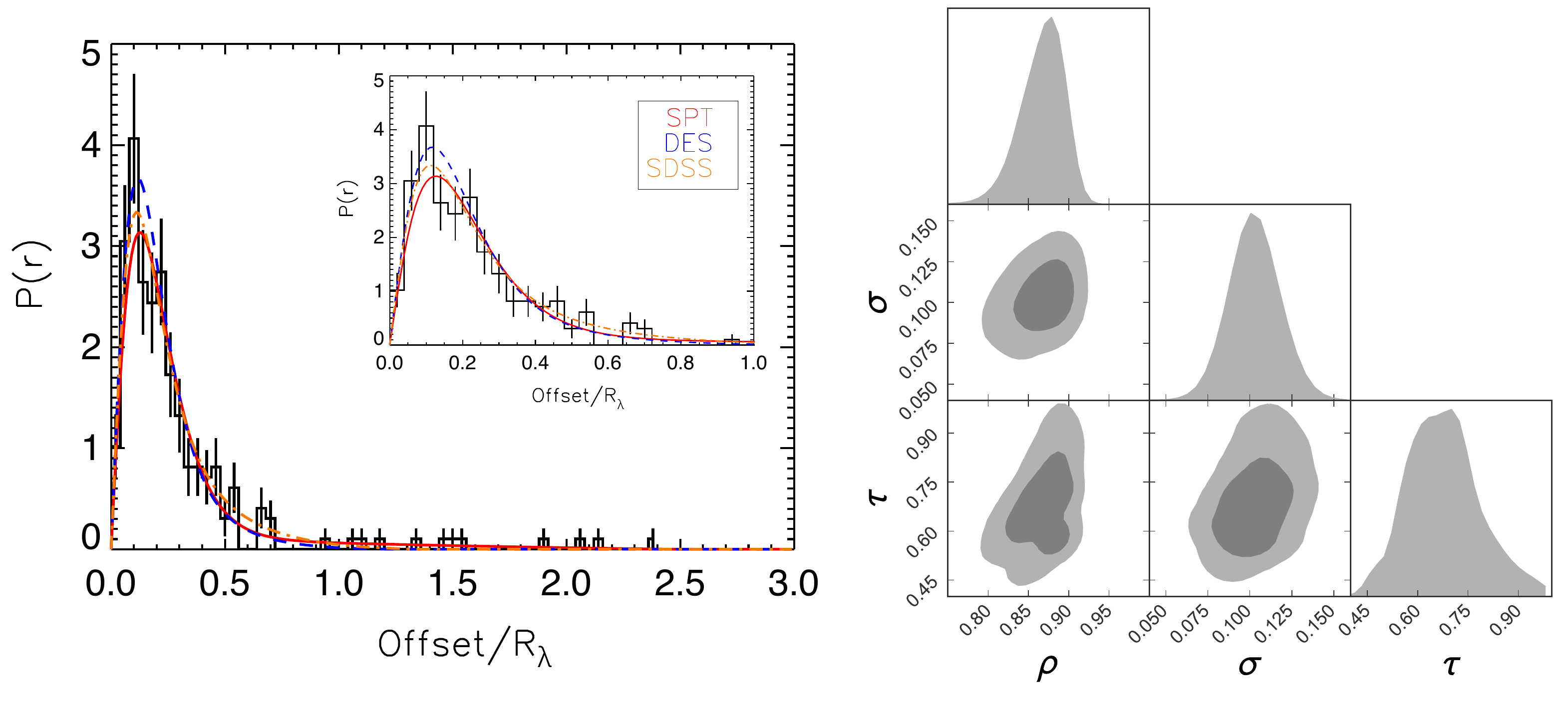}
\caption{(Left) The distribution of separations between SPT centers and RM most probable central galaxies as a fraction of the RM cluster radius $R_\lambda$ for systems at $\lambda>20$, the richness threshold for DES cosmological analyses. Overplotted in red is the best fit to the SPT data and in orange (blue) are the best-fit models  from \citet{zhang19} convolved with the SPT positional uncertainty. This latter analysis characterized the offsets between X-ray peaks and RM central galaxies  for 144(67) systems in SDSS (DES). The large SPT-RM sample shows a higher fraction with large offsets than previous works.  (Right) The best-fit model constraints. These results are overplotted in red in the left panel. We note that the derived value of $\tau$ is very sensitive to clusters with large separation.} 
\label{fig:centering2}
\end{center}
\end{figure*}

As a second model of the SZ-central galaxy offset distribution we  explore a miscentering model tied to the RM cluster radius, $R_\lambda$, where 
\begin{equation}
R_\lambda = (\frac{\lambda}{100})^{0.2} h^{-1} \textrm{ \  Mpc.}
\end{equation} 
$R_\lambda$ is determined by the RM cluster finding algorithm and corresponds to the maximum separation between the RM central galaxy and cluster members that contribute to the optical richness measurement. 
Here we focus on the better constrained SPT clusters at $\xi>5$ and we plot this distribution in Figure \ref{fig:centering2}.  As can be seen, there are a significant number of systems (14 of 249, $6\%$) that have offsets greater than $R_\lambda$. 
Examination of these clusters shows, unsurprisingly, that they display similar characteristics to the outliers in the previous subsection (and many are in common). 
Additionally it is worth noting that issues that reduce the richness estimate will more adversely affect a fractional offset when the cluster scale is set by the richness measure (e.g., $R_\lambda$) as opposed to being set by the SZ mass estimate.

Following  \citet{mcclintock19} and \citet{zhang19}, we model the probability distribution for the separation between SZ centroids and RM central galaxies as the combination of an exponential distribution that reflects the well-centered systems and a Gamma distribution $\Gamma(2, \tau)$ that characterizes those clusters with larger separations: 

\begin{equation}\label{eq:pofx2}
P(x) = \rho \frac{1}{\sigma_\lambda}e^{-\frac{x}{\sigma_\lambda}} + (1-\rho)\frac{x}{\tau^2}e^{-\frac{x}{\tau}}
\end{equation}
where now $x=r_\textrm{offset}/R_\lambda$, $\sigma_\lambda$ characterizes the exponential distribution, $\tau$ is the scale parameter of the Gamma distribution function and, as in the previous model, we also incorporate the SPT positional uncertainties when conducting the fit.  

The two-dimensional convolutions required for properly incorporating the SPT positional uncertainty in this model are computationally expensive to repeat many times in an  MCMC analysis. 
Numerical computations of the probability distribution can instead be replaced by relatively inexpensive, yet highly precise emulators. 
For this purpose, we use Gaussian Processes (GP, \citealt{rasmussen06}), a method that has facilitated robust forward modeling of various astrophysical functions (e.g., \citealt{heitmann06}, \citealt{habib07} and other applications). 
We detail the construction and validation of our emulator of the miscentering distribution model in Appendix \ref{sec:emulator}. 

For this analysis we adopt the priors: 
\begin{subequations}
\begin{align*}
		0.3 \leq  & \rho \leq  1.0 \\
		0.001 \leq & \sigma_\lambda \leq  0.25 \\ 
		0.05 \leq & \tau \leq  1.0
\end{align*}
\end{subequations}

We plot these results in Figure \ref{fig:centering2} and report the parameter constraints in Table \ref{tab:offset_model2tab}. 
In Figure \ref{fig:centering2} we also over plot the best-fit model curves from \citet{zhang19} convolved with the SPT positional uncertainty. 

While \citet{zhang19} explored the separation between X-ray peaks and central galaxies,  the analysis here quantifies the central galaxy offset from the gas center averaged over a larger scale via the SPT matched filter. This should generally have a small effect;  studies with X-ray centering proxies  (see e..g, \citealt{mann12}) have found an additional 20-60 kpc ($\sim0.02-0.06R_\lambda$) scatter in the BCG and X-ray centroid separation (to which our measurement is most analogous) as compared to the X-ray peak to BCG separation though there can be notable outliers in the case of merging clusters. 
With this caveat in mind,  we find that our results at $\xi>5$, with $\rho=0.87^{+0.02}_{-0.03}$ of the clusters within the ``well-centered'' component of the distribution agree with previous RM results on DES ($\rho=0.84^{+0.11}_{-0.07}$) and are higher than those found in SDSS ($\rho=0.68^{+0.03}_{-0.05}$).
However, our recovered value of  $\tau$ is notably higher than previous results ($\tau=0.69^{+0.12}_{-0.09}$, versus $0.16^{+0.11}_{-0.04}$) as it is significantly affected by clusters in the long tail. In comparison \citet{zhang19}  only found 1 of 67 systems (1.5\%) with gas-BCG separations at $R > R_\lambda$ compared to the 14 (6\%) found here.  If we reanalyze the cluster sample excluding systems with offsets $R>R_\lambda$, we find  $\rho$ and $\sigma_\lambda$ significantly less well constrained ($\rho=0.74^{+0.22}_{-0.30}$, $\sigma_\lambda=0.105^{+0.045}_{-0.07}$) and $\tau$ shifted to lower values consistent with previous work ($\tau=0.13^{+0.075}_{-0.045}$).
It will be important in future weak lensing analyses to quantify this tail while incorporating all the cluster selection effects relevant to the analysis at hand as \citet{zhang19} found that shifts in $\tau$ at the 0.04 level can lead to systematic shifts in the weak lensing derived mass calibrations at the level of $\delta\textrm{log}M_{200}= 0.015.$

\begin{deluxetable}{ccc}[t]  
\tablecolumns{2}
\tablecaption{\\ Miscentering Model 2 Fits \label{tab:offset_model2tab}}
\tablehead{ 
\colhead{Parameter} &
\colhead{$\xi \ge 5$} & 
 } 
 \startdata 
$\hspace{0.5cm} \rho   \hspace{.1cm} $ & $\hspace{1cm} 0.87^{+0.02}_{-0.03}$\\
$\hspace{0.5cm} \sigma \hspace{.1cm}  $ & $\hspace{1cm} 0.12^{+0.015}_{-0.01} $ \\ 
$\hspace{0.5cm} \tau    \hspace{.1cm} $ & $\hspace{1cm} 0.69^{+0.12}_{-0.09} $
\enddata 
\tablecomments{Best-fit miscentering parameters for the SPT-RM Volume Limited Sample as characterized in Equation \ref{eq:pofx2}. }
\end{deluxetable}

\section{Conclusions}
In this work, we describe the \surveyname, a new 2770 deg$^2$ survey conducted at 95 and 150 GHz using the \sptpol\ receiver. 
Using a matched spatial-spectral filter with a SZ detection significance threshold of $\xi \ge 5$, we have identified \ncandfive\ cluster candidates.  
Of these, we have confirmed and estimated redshifts for \nconfirmfive\ clusters 
using a combination of external optical imaging data, primarily from the DES survey, and targeted observations with the Magellan/PISCO imager. 
With more incomplete followup, we also confirm an additional \nconfirmfourtofive \ systems at $4<\xi<5$.  Approximately two-thirds of the confirmed clusters are first reported in this work. 

We estimate cluster masses using a $\xi$-mass scaling relation, inferred from fitting the observed SZ-cluster density at $\xi>5$ and redshift $z>0.25$ to a fixed spatially flat $\Lambda$CDM cosmology. The \surveyshort\ cluster sample has a median redshift of {$z=0.49$} with 20 clusters at $z > 1$, a median mass of $M_{500c} \sim {4.4 \times 10^{14} M_\odot h^{-1}}$, and we unambiguously identify strong gravitational lensing in \nstrong \ systems.  
Selected data products for this catalog will be hosted at \webaddress. 

We use 1.4 GHz observations from NVSS to estimate the amount of radio contamination in the \surveyshort\ sample.  We estimate a median radio contamination of 0.05 in units of the SZ detection significance, which is $\sim$1\% of the SZ signal at the $\xi=5$ detection threshold.  We find that only $\sim$5\% of these candidates would have a predicted radio contamination of $>$10\% compared to the SZ signal level. When extending this test to consider only confirmed clusters at $\xi \gtrsim 4$, we find $<4$\% of these clusters would have a predicted radio contamination of $>$10\% compared to the SZ signal level.
As this test was performed using an SZ-selected sample, it places a lower limit to the radio contamination of the SZ signal of massive clusters, as clusters with extremely bright radio sources could be missed by our SZ selection altogether.  However, as discussed in Sections \ref{sec:contam} and \ref{sec:maskedregion}, such occurrences are expected to be rare  at the redshifts of most interest for the SPT sample ($z>0.25$). 

We next associate SZ-selected cluster candidates from a combination of the \surveyshort\ and SPT-SZ surveys with clusters from both the \textit{Planck} PSZ2 sample and the DES Year 3 redMaPPer cluster catalog. 
We find general agreement with previous studies assessing the consistency of \textit{Planck}- and SPT-derived masses, and that, as expected, the SPT catalogs contain the majority of PSZ2 clusters at $z>0.25$ in the SPT footprint.

Considering the SPT and DES RM catalogs, we find \nredmapperspt \  clusters that match with a false association probability $<5\%$ at $\xi>4.5$.  When restricting this comparison to the redMaPPer volume-limited catalog at $z > 0.25$, we identify  \nredmapperzcutvol \  systems.  Using this sample, we characterize the offset distribution between the SZ center and central galaxy.  We  find general agreement with the constraints from previous studies \citep{saro15, zhang19} but note our large sample size allows us to identify a significant tail of clusters to large separations not present in these previous works.  We also use the SZ-mass estimates to constrain the optical richness-mass relation assuming a fixed standard cosmology.  We find that our relation intersects with the previous weak lensing studies of \citet{mcclintock19,raghunathan19a} at a richness of $\lambda=60$, but that our SPT derived relation prefers a 28\% shallower slope with the difference significant at the $4.0 \sigma$ level.
To reproduce the slope of the weak lensing analysis we would require a significant shift in our assumed cosmology, but we leave any quantitative conclusions to a future analysis.  Regardless, our work highlights the value of consistency checks between scaling relations inferred from multi-wavelength observations, which should lead to constraints with better understood systematic uncertainties.

Combined with clusters detected from the SPT-SZ (B15) and SPTpol 100d surveys \citep{huang19}, this work increases the number of SZ-detected clusters reported by the South Pole Telescope to more than 1,000.
Future SZ-selected cluster catalogs from the SPT will push to higher redshift and lower mass.  From \sptpol, this includes the catalog from the completed 500 deg$^2$ survey, which is a factor of 5-10 deeper than  \surveyshort \ \citep{henning18}.  The ongoing 1500 deg$^2$ SPT-3G survey \citep{benson14} is expected to be even deeper, with a mass-selection threshold of $\sim 10^{14} M_\odot h^{-1}$, which will  enable the detection of $\gtrsim 4000$ clusters. This work will complement the wide-area cluster surveys to be conducted at X-ray (\erosita, \citealt{predehl10, merloni12}) and optical/IR wavelengths (e.g., LSST, \citealt{lsst09}, \textit{Euclid}, \citealt{euclid19}, \textit{WFIRST}, \citealt{spergel15}), as well as SZ surveys by AdvACT \citep{henderson16} and Simons Observatory \citep{ade19}, with all of the SZ surveys ultimately setting the stage for the next-generation CMB-S4 survey \citep{abazajian16, abazajian19}.  

\section*{Acknowledgements} 

\facilities{Blanco (DECAM), Hubble Space Telescope (WFC3), Magellan:Baade (FourStar), Magellan:Clay (PISCO, LDSS3C), NSF/US Department of Energy 10m South Pole Telescope (SPTpol), Spitzer (IRAC), AAT(2dF+AAOmega)}

\smallskip

This work was performed in the context of the SouthPole Telescope scientific program. SPT is supported by the National Science Foundation through grant PLR-1248097. Partial support is also provided by the NSF Physics Frontier Center grant PHY-0114422 to the Kavli Institute of Cosmological Physics at the University of Chicago, the Kavli Foundation and the Gordon and Betty Moore Foundation grant GBMF 947 to  the  University  of  Chicago. This  work is  also  supported  by  the  U.S.  Department  of  Energy.  PISCO observations are supported by NSF AST-1814719.

Work at Argonne National Lab is supported by UChicago Argonne LLC,Operator  of  Argonne  National  Laboratory  (Argonne). Argonne, a U.S. Department of Energy Office of Science Laboratory,  is  operated  under  contract  no.   DE-AC02-06CH11357.  We also acknowledge support from the Argonne  Center  for  Nanoscale  Materials. 
MG and LB acknowledge partial support from HST-GO-15307.001.
BB is supported by the FermiResearch  Alliance  LLC  under  contract  no.   De-AC02-07CH11359  with  the  U.S.  Department  of  Energy.    The CU Boulder group acknowledges support from NSF AST-0956135.  The McGill authors acknowledge funding from the Natural Sciences and Engineering Research Council of  Canada, Canadian  Institute  for  Advanced  Research, and the Fonds de Recherche du Qu\'{e}bec Nature et technologies.
The UCLA authors acknowledge support from NSF AST-1716965 and CSSI-1835865.  The Stanford/SLAC group acknowledges support from the U.S. Department of Energy under contract number DE-AC02-76SF00515.
AS is supported by the ERC-StG `ClustersXCosmo' grant agreement 716762, and by the FARE-MIUR grant `ClustersXEuclid' R165SBKTMA.
CH acknowledges support from the Max Planck Society and the Alexander von Humboldt Foundation, in the framework of the Max Planck-Humboldt Research Award endowed by the Federal Ministry of Education and Research, in addition to support from the European Research Council under grant number 647112. 
SJ acknowledges support from the Beecroft Trust and ERC 693024.
TS acknowledges support from the German Federal Ministry of Economics and Technology (BMWi) provided through DLR under projects 50 OR 1610 and 50 OR 1803, as well as support from the Deutsche Forschungsgemeinschaft, DFG, under project SCHR 1400/3-1.
The Melbourne authors acknowledge support from the Australian Research Council's Discovery Projects scheme (DP150103208).

The 2dFLenS survey is based on data acquired through the Australian Astronomical Observatory, under program A/2014B/008.
This work is based in part on observations made with the Spitzer Space Telescope, which is operated by the Jet Propulsion Laboratory, California Institute of Technology under a contract with NASA.

Funding for the DES Projects has been provided by the U.S. Department of Energy, the U.S. National Science Foundation, the Ministry of Science and Education of Spain, 
the Science and Technology Facilities Council of the United Kingdom, the Higher Education Funding Council for England, the National Center for Supercomputing 
Applications at the University of Illinois at Urbana-Champaign, the Kavli Institute of Cosmological Physics at the University of Chicago, 
the Center for Cosmology and Astro-Particle Physics at the Ohio State University,
the Mitchell Institute for Fundamental Physics and Astronomy at Texas A\&M University, Financiadora de Estudos e Projetos, 
Funda{\c c}{\~a}o Carlos Chagas Filho de Amparo {\`a} Pesquisa do Estado do Rio de Janeiro, Conselho Nacional de Desenvolvimento Cient{\'i}fico e Tecnol{\'o}gico and 
the Minist{\'e}rio da Ci{\^e}ncia, Tecnologia e Inova{\c c}{\~a}o, the Deutsche Forschungsgemeinschaft and the Collaborating Institutions in the Dark Energy Survey. 

The Collaborating Institutions are Argonne National Laboratory, the University of California at Santa Cruz, the University of Cambridge, Centro de Investigaciones Energ{\'e}ticas, 
Medioambientales y Tecnol{\'o}gicas-Madrid, the University of Chicago, University College London, the DES-Brazil Consortium, the University of Edinburgh, 
the Eidgen{\"o}ssische Technische Hochschule (ETH) Z{\"u}rich, 
Fermi National Accelerator Laboratory, the University of Illinois at Urbana-Champaign, the Institut de Ci{\`e}ncies de l'Espai (IEEC/CSIC), 
the Institut de F{\'i}sica d'Altes Energies, Lawrence Berkeley National Laboratory, the Ludwig-Maximilians Universit{\"a}t M{\"u}nchen and the associated Excellence Cluster Universe, 
the University of Michigan, the National Optical Astronomy Observatory, the University of Nottingham, The Ohio State University, the University of Pennsylvania, the University of Portsmouth, 
SLAC National Accelerator Laboratory, Stanford University, the University of Sussex, Texas A\&M University, and the OzDES Membership Consortium.

Based in part on observations at Cerro Tololo Inter-American Observatory, National Optical Astronomy Observatory, which is operated by the Association of 
Universities for Research in Astronomy (AURA) under a cooperative agreement with the National Science Foundation.

The DES data management system is supported by the National Science Foundation under Grant Numbers AST-1138766 and AST-1536171.
The DES participants from Spanish institutions are partially supported by MINECO under grants AYA2015-71825, ESP2015-66861, FPA2015-68048, SEV-2016-0588, SEV-2016-0597, and MDM-2015-0509, 
some of which include ERDF funds from the European Union. IFAE is partially funded by the CERCA program of the Generalitat de Catalunya.
Research leading to these results has received funding from the European Research
Council under the European Union's Seventh Framework Program (FP7/2007-2013) including ERC grant agreements 240672, 291329, and 306478.
We  acknowledge support from the Brazilian Instituto Nacional de Ci\^encia
e Tecnologia (INCT) e-Universe (CNPq grant 465376/2014-2).

This manuscript has been authored by Fermi Research Alliance, LLC under Contract No. DE-AC02-07CH11359 with the U.S. Department of Energy, Office of Science, Office of High Energy Physics. The United States Government retains and the publisher, by accepting the article for publication, acknowledges that the United States Government retains a non-exclusive, paid-up, irrevocable, world-wide license to publish or reproduce the published form of this manuscript, or allow others to do so, for United States Government purposes.

The Pan-STARRS1 Surveys (PS1) and the PS1 public science archive have been made possible through contributions by the Institute for Astronomy, the University of Hawaii, the Pan-STARRS Project Office, the Max-Planck Society and its participating institutes, the Max Planck Institute for Astronomy, Heidelberg and the Max Planck Institute for Extraterrestrial Physics, Garching, The Johns Hopkins University, Durham University, the University of Edinburgh, the Queen's University Belfast, the Harvard-Smithsonian Center for Astrophysics, the Las Cumbres Observatory Global Telescope Network Incorporated, the National Central University of Taiwan, the Space Telescope Science Institute, the National Aeronautics and Space Administration under Grant No. NNX08AR22G issued through the Planetary Science Division of the NASA Science Mission Directorate, the National Science Foundation Grant No. AST-1238877, the University of Maryland, Eotvos Lorand University (ELTE), the Los Alamos National Laboratory, and the Gordon and Betty Moore Foundation.

\bibliography{../../BIBTEX/spt}

\appendix

\section{Offset Distribution Emulator Construction}\label{sec:emulator}

In this section we describe the construction of our emulator of the SZ - RM central galaxy offset distribution discussed in Section \ref{subsec:descentering}. This distribution is modeled as the SPT positional uncertainty (see Eq'n \ref{eq:sptpos}) convolved with: 
\begin{equation}
P(x) = \rho \frac{1}{\sigma_\lambda}e^{-\frac{x}{\sigma_\lambda}} + (1-\rho)\frac{x}{\tau^2}e^{-\frac{x}{\tau}}
\end{equation}
where $x=r_\textrm{offset}/R_\lambda$, $\sigma_\lambda$ characterizes the exponential distribution, and $\tau$ is the scale parameter of the Gamma distribution function. 
The training probability distributions $p(x, \theta)$ are generated at $N=1024$ points on a latin hypercube sampling (LHS) design of the 3 centering model parameters as well as the SPT positional uncertainty scaled by the RM size (converted to radians) $\theta = \{\rho, \sigma_\lambda, \tau, \sigma_{SPT}/\theta_\lambda \}$. 
As shown in Cosmic Emulators \citep{heitmann16}, the space-filling properties of LHS are well-suited for GP interpolation on a relatively small number of training points. The range of our centering model parameters are identical to the flat priors in our likelihood analysis and the SPT positional uncertainty trained over the range $0.0 \leq  \sigma_{SPT}/\theta_\lambda \leq  1.0$.

Our emulation strategy also follows that of the Cosmic Emulators. 
That is, we first perform a singular value decomposition of probability $p(x, \theta)$ values in $100$ bins (spanning separations from 0 to $R_\lambda$). 
Weights of $16$ truncated orthogonal bases are then modeled as independent functions of input parameters $\{ \rho, \sigma_\lambda, \tau, \sigma_{SPT}/R_\lambda \}$ using GP as a local interpolating scheme. 
The key ingredient of {\it learning} in GP modeling is the configuration of the covariance function and determination of the associated hyperparameters, which we find using Bayesian optimization. 
We also check the robustness of the emulator accuracy with different choices of covariance functions. 

The fully trained emulator is validated on the parameter values within the limits of the latin hypercube, but not at the specific points where the emulator is fitted. 
The evaluation time for the trained emulator is less than $0.001$ seconds per computation, delivering a speed-up of $1000$ over numerical calculation of $p(x, \theta)$. 
This is crucial for quick explorations of the posterior distribution of parameters, where our GP emulator is implemented in the MCMC likelihood calculation. 

\clearpage

\section{Associations with LRGs from the 2dFLenS Survey}
SPT clusters, redshift estimates, and associated LRG spectroscopic redshifts for clusters associated with LRGs in the 2dFLenS Survey (see Section \ref{sec:tdfsec}). SPT-CL~J0302$-$3306 and SPT-CL~J0319$-$2853 were targeted in a spare fiber program. The full 2dFLenS redshift catalog is available at \url{http://2dflens.swin.edu.au/}.
\tabletypesize{\footnotesize}
\begin{deluxetable}{lccc}[h]  
\tablecolumns{4}
\tablecaption{\\ Associations with 2dFLenS LRGs\label{tab:tdf_match}}
\tablehead{ 
\colhead{SPT ID} &
\colhead{$z$} & 
\colhead{2dFLenS $z$(s)} &
\colhead{Offset (arcmin)}}
\startdata
SPT$-$CL J0000$-$2805 & $0.23\pm 0.03 $ & $ 0.283 $ & $ 0.64$\\
SPT$-$CL J0014$-$3022 & $0.307 $ & $ 0.308, 0.317 $ & $ 1.04, 1.68$\\
SPT$-$CL J0036$-$3144 & $0.41\pm 0.01 $ & $ 0.413 $ & $ 0.18$\\
SPT$-$CL J0042$-$2831 & $0.109 $ & $ 0.110, 0.109 $ & $ 1.16, 2.46$\\
SPT$-$CL J0100$-$3246 & $0.53\pm 0.01 $ & $ 0.532 $ & $ 1.76$\\
SPT$-$CL J0114$-$2820 & $0.43\pm 0.01 $ & $ 0.441, 0.447 $ & $ 1.03, 1.22$\\
SPT$-$CL J0115$-$2917 & $0.41\pm 0.01 $ & $ 0.397 $ & $ 1.27$\\
SPT$-$CL J0121$-$3355 & $0.57\pm 0.01 $ & $ 0.579 $ & $ 1.28$\\
SPT$-$CL J0152$-$2853 & $0.413 $ & $ 0.416, 0.406 $ & $ 0.25, 0.85$\\
SPT$-$CL J0158$-$2910 & $0.57\pm 0.01 $ & $ 0.576 $ & $ 2.11$\\
SPT$-$CL J0159$-$3010 & $0.69\pm 0.01 $ & $ 0.699, 0.703 $ & $ 1.21, 1.57$\\
SPT$-$CL J0159$-$3331 & $0.40\pm 0.01 $ & $ 0.411, 0.406 $ & $ 0.42, 2.31$\\
SPT$-$CL J0202$-$2812 & $0.12\pm 0.01 $ & $ 0.111 $ & $ 2.32$\\
SPT$-$CL J0202$-$3027 & $0.48\pm 0.01 $ & $ 0.489, 0.493 $ & $ 0.27, 1.52$\\
SPT$-$CL J0206$-$2921 & $0.28\pm 0.01 $ & $ 0.273 $ & $ 1.60$\\
SPT$-$CL J0215$-$2948 & $0.25\pm 0.01 $ & $ 0.256 $ & $ 2.36$\\
SPT$-$CL J0217$-$3200 & $0.35\pm 0.01 $ & $ 0.341 $ & $ 0.26$\\
SPT$-$CL J0218$-$3142 & $0.27\pm 0.01 $ & $ 0.275, 0.269, 0.267 $ & $ 0.19, 0.74, 1.68$\\
SPT$-$CL J0224$-$3223 & $0.54\pm 0.01 $ & $ 0.545, 0.545 $ & $ 1.36, 1.86$\\
SPT$-$CL J0241$-$2839 & $0.238 $ & $ 0.226, 0.237 $ & $ 0.89, 2.45$\\
SPT$-$CL J0242$-$3123 & $0.50\pm 0.01 $ & $ 0.491 $ & $ 1.01$\\
SPT$-$CL J0302$-$3209 & $0.32\pm 0.01 $ & $ 0.327, 0.325 $ & $ 0.18, 1.67$\\
SPT$-$CL J0302$-$3306 & $0.73\pm 0.01 $ & $ 0.752 $ & $ 0.66$\\
SPT$-$CL J0303$-$2736 & $0.27\pm 0.01 $ & $ 0.261 $ & $ 0.73$\\
SPT$-$CL J0305$-$3229 & $0.53\pm 0.01 $ & $ 0.529 $ & $ 0.33$\\
SPT$-$CL J0307$-$2840 & $0.253 $ & $ 0.250 $ & $ 0.45$\\
SPT$-$CL J0309$-$3209 & $0.54\pm 0.01 $ & $ 0.526 $ & $ 2.35$\\
SPT$-$CL J0319$-$2853 & $0.36\pm 0.01 $ & $ 0.355 $ & $ 0.04$\\
SPT$-$CL J0319$-$3345 & $0.41\pm 0.01 $ & $ 0.411 $ & $ 0.46$\\
SPT$-$CL J2159$-$2846 & $0.43\pm 0.04 $ & $ 0.423, 0.431 $ & $ 0.89, 2.19$\\
SPT$-$CL J2220$-$3509 & $0.154 $ & $ 0.152 $ & $ 1.55$\\
SPT$-$CL J2234$-$3033 & $0.251 $ & $ 0.246 $ & $ 0.73$\\
SPT$-$CL J2234$-$3159 & $0.57\pm 0.04 $ & $ 0.557 $ & $ 0.46$\\
SPT$-$CL J2251$-$3324 & $0.24\pm 0.02 $ & $ 0.230, 0.231 $ & $ 0.43, 1.00$\\
SPT$-$CL J2253$-$3344 & $0.224 $ & $ 0.228 $ & $ 0.07$\\
SPT$-$CL J2258$-$3447 & $0.317 $ & $ 0.307, 0.308 $ & $ 0.64, 0.97$\\
SPT$-$CL J2321$-$2725 & $0.67\pm 0.04 $ & $ 0.658 $ & $ 0.92$\\
SPT$-$CL J2335$-$3256 & $0.51\pm 0.04 $ & $ 0.490, 0.511 $ & $ 1.91, 2.16$\\
SPT$-$CL J2336$-$3205 & $0.63\pm 0.04 $ & $ 0.619, 0.613, 0.623 $ & $ 0.27, 1.52, 2.43$

\enddata 
\tablecomments{For each cluster we report the cluster name, spectroscopic (3 digits) or photometric  (2 digits with uncertainty) redshift, the spectroscopic redshifts of the 2dFLenS LRGS, and the spatial separation of these LRGs from the SPT cluster location.}
\end{deluxetable}
\clearpage 

\section{The Cluster Catalogs}
In this section we provide three different tables: the complete cluster candidate list at $\xi>5$ from \surveyshort, the confirmed sample from \surveyshort \ at $4<\xi<5$, and, finally, newly-confirmed clusters at $\xi>4.5$  from the 2500d SPT-SZ survey \citep{bleem15b}. The SPT-SZ clusters were confirmed using our RM association process described in Section \ref{subsec:desmatch}. The data from these tables, including references for the sources of spectroscopic redshifts, photometric redshifts(when taken from the literature), and strong lensing information (where previously known), as well as additional notes on individual clusters are available online at  \webaddress. 

\startlongtable
\begin{center}
\begin{deluxetable*}{ lcc | cc |c|c|c|c|c}
\tablecaption{Galaxy cluster candidates above $\xi = 5$ in the \surveyname.
\label{tab:sampletable}}
\tablehead{
\multicolumn{1}{c}{\bf SPT ID} & 
\multicolumn{1}{c}{\bf R.A.} & 
\multicolumn{1}{c}{\bf Decl.} & 
\multicolumn{2}{c}{\bf Best} &
\multicolumn{1}{c}{\bf Redshift} &
\multicolumn{1}{c}{\bf M$_{500c}$}  &
\multicolumn{1}{c}{\bf $\lambda$}  &
\multicolumn{1}{c}{\bf Image Source}  &
\multicolumn{1}{c}{\bf Notes} \\
\colhead{} & 
\colhead{(J2000)} & 
\colhead{(J2000)}  & 
\colhead{\nSN{}} & 
\colhead{$\theta_\textrm{c}$} & 
\colhead{} &
\colhead{($10^{14} h_{70}^{-1} M_\odot$)} &
\colhead{}  &
\colhead{} &
\colhead{}
}
\startdata 
SPT-CL J0000$-$2518 & 0.0723 & $-25.3039$ & 5.07 & 1.50 & - & - & - & 1 & -\\
SPT-CL J0005$-$3751 & 1.4383 & $-37.8544$ & 6.34 & 0.75 & 0.483$\pm$0.007 & $5.42^{+0.89}_{-0.89}$ & 86$\pm$5 & 3 & SL\\
SPT-CL J0005$-$3443 & 1.4877 & $-34.7193$ & 5.51 & 2.00 & 0.114 & $5.32^{+0.90}_{-1.00}$ & 105$\pm$5 & 8 & -\\
SPT-CL J0012$-$3537 & 3.0433 & $-35.6226$ & 5.26 & 2.25 & 0.69$\pm$0.02 & $4.38^{+0.80}_{-0.89}$ & 73$\pm$5 & 3 & -\\
SPT-CL J0012$-$3346 & 3.1112 & $-33.7782$ & 6.70 & 1.00 & 0.68$\pm$0.01 & $5.37^{+0.82}_{-0.85}$ & 101$\pm$6 & 3 & -\\
SPT-CL J0014$-$3022 & 3.5727 & $-30.3831$ & 18.29 & 1.50 & 0.307 & $11.39^{+1.15}_{-1.38}$ & - & 8 & SL\\
SPT-CL J0014$-$2024 & 3.6932 & $-20.4087$ & 5.64 & 1.00 & 0.317 & $5.62^{+0.98}_{-0.98}$ & - & 8 & -\\
SPT-CL J0020$-$2543 & 5.1731 & $-25.7276$ & 7.14 & 1.50 & 0.141 & $7.04^{+0.97}_{-1.08}$ & 101$\pm$18 & 8 & -\\
\enddata
\tablecomments{Here we report for each cluster candidate detected at $\xi>5$ the candidate name, position, detection significance ($\xi$), the $\beta$-model core radius (in arcminutes) corresponding to this significance (see Section \ref{sec:extract}), and the source of followup imaging. When a candidate is confirmed as a cluster we also report its redshift and mass, richness from RM in ``scanning mode''' (where available, see Section \ref{subsec:desmatch}), and flag if the system has been identified as a strong gravitational lens (SL) or if the estimated radio contamination to the SZ signal exceeds 10\% (RC; see Section \ref{sec:contam}). The key for the Image Source column is as follows: [1] Pan-STARRS, [2]Magellan/PISCO, [3] DES, [4] Magellan/FourStar, [5] \textit{Spitzer}/IRAC, [6] WISE, [7] Literature photometric redshift, [8] Spectroscopic redshift.  Generally, data from Pan-STARRS  is deep enough to confirm clusters to $z\sim0.6$, DES and PISCO to $z\sim0.8-1.0$,  FourStar to $z\sim1.2$ and  \textit{Spitzer}  to $z\sim1.5$. }
\end{deluxetable*}

\clearpage

\startlongtable
\begin{deluxetable*}{ lcc | cc |c|c|c|c|c}
\tablecaption{Confirmed galaxy clusters  $4< \xi < 5$ in the \surveyname. 
\label{tab:sample4sigma}}
\tablehead{
\multicolumn{1}{c}{\bf SPT ID} & 
\multicolumn{1}{c}{\bf R.A.} & 
\multicolumn{1}{c}{\bf Decl.} & 
\multicolumn{2}{c}{\bf Best} &
\multicolumn{1}{c}{\bf Redshift} &
\multicolumn{1}{c}{\bf M$_{500c}$}  &
\multicolumn{1}{c}{\bf $\lambda$}  &
\multicolumn{1}{c}{\bf Imaging}  &
\multicolumn{1}{c}{\bf Notes} \\
\colhead{} & 
\colhead{(J2000)} & 
\colhead{(J2000)}  & 
\colhead{\nSN{}} & 
\colhead{$\theta_\textrm{c}$} & 
\colhead{} &
\colhead{($10^{14} h_{70}^{-1} M_\odot$)} &
\colhead{}  &
\colhead{} &
\colhead{}
}
\startdata 
SPT-CL J0000$-$3838 & 0.0118 & $-38.6467$ & 4.37 & 2.25 & 0.305$\pm$0.007 & $4.25^{+0.71}_{-0.95}$ & 49$\pm$3 & 3 & -\\
SPT-CL J0000$-$2805 & 0.0906 & $-28.0947$ & 4.79 & 2.50 & 0.23$\pm$0.03 & $5.05^{+0.91}_{-1.09}$ & - & 1 & -\\
SPT-CL J0001$-$3446 & 0.3080 & $-34.7729$ & 4.23 & 0.25 & 0.73$\pm$0.02 & $3.72^{+0.57}_{-0.83}$ & 45$\pm$7 & 3 & -\\
SPT-CL J0011$-$2841 & 2.8272 & $-28.6859$ & 4.13 & 0.25 & 0.089$\pm$0.006 & $4.79^{+0.74}_{-1.05}$ & 22$\pm$5 & 3 & -\\
SPT-CL J0019$-$2026 & 4.7834 & $-20.4475$ & 4.95 & 1.25 & 0.277 & $5.12^{+0.98}_{-1.05}$ & - & 8 & -\\
SPT-CL J0020$-$2634 & 5.2048 & $-26.5799$ & 4.14 & 0.25 & 0.233$\pm$0.006 & $4.62^{+0.74}_{-1.02}$ & 22$\pm$4 & 3 & -\\
SPT-CL J0023$-$3252 & 5.7941 & $-32.8672$ & 4.28 & 0.75 & 0.66$\pm$0.01 & $3.83^{+0.61}_{-0.84}$ & 45$\pm$4 & 3 & -\\
SPT-CL J0025$-$3618 & 6.3664 & $-36.3085$ & 4.53 & 0.25 & 0.215$\pm$0.005 & $4.48^{+0.77}_{-0.99}$ & 21$\pm$2 & 3 & -\\
\enddata
\tablecomments{The same as Table \ref{tab:sampletable} now for confirmed candidates at $4 < \xi<5$.}
\end{deluxetable*}
\end{center}

\clearpage

\begin{center}
\begin{deluxetable*}{ lcc | cc |c|c|c}[h]
\tablecaption{Newly confirmed galaxy clusters at $\xi > 4.5$ in the SPT-SZ Survey.
\label{tab:newsptsz}}
\tablehead{
\multicolumn{1}{c}{\bf SPT ID} & 
\multicolumn{1}{c}{\bf R.A.} & 
\multicolumn{1}{c}{\bf Decl.} & 
\multicolumn{2}{c}{\bf Best} &
\multicolumn{1}{c}{\bf Redshift} &
\multicolumn{1}{c}{\bf M$_{500c}$}  &
\multicolumn{1}{c}{\bf $\lambda$}  \\
\colhead{} & 
\colhead{(J2000)} & 
\colhead{(J2000)}  & 
\colhead{\nSN{}} & 
\colhead{$\theta_\textrm{c}$} & 
\colhead{} &
\colhead{($10^{14} h_{70}^{-1} M_\odot$)} &
\colhead{}  
}
\startdata 
SPT-CL J2352$-$5846 & 358.0510 & $-58.7758$ & 5.18 & 0.75 & 0.167$\pm$0.006 & $3.86^{+0.67}_{-0.77}$ & 23$\pm$6\\
SPT-CL J0013$-$5714 & 3.3029 & $-57.2373$ & 5.11 & 1.50 & 0.63$\pm$0.01 & $3.70^{+0.67}_{-0.76}$ & 29$\pm$3\\
SPT-CL J2328$-$4616 & 352.0576 & $-46.2802$ & 4.84 & 2.75 & 0.224$\pm$0.005 & $3.94^{+0.70}_{-0.84}$ & 25$\pm$3\\
SPT-CL J0048$-$4450 & 12.1743 & $-44.8475$ & 4.82 & 0.50 & 0.595$\pm$0.009 & $3.55^{+0.61}_{-0.77}$ & 55$\pm$5\\
SPT-CL J2339$-$4058 & 354.7996 & $-40.9697$ & 4.75 & 0.25 & 0.60$\pm$0.01 & $3.50^{+0.61}_{-0.76}$ & 31$\pm$3\\
SPT-CL J2136$-$5723 & 324.1203 & $-57.3968$ & 4.72 & 1.00 & 0.284$\pm$0.007 & $3.65^{+0.63}_{-0.78}$ & 42$\pm$3\\
SPT-CL J2158$-$4851 & 329.5692 & $-48.8533$ & 4.64 & 0.25 & 0.490$\pm$0.008 & $3.69^{+0.65}_{-0.82}$ & 58$\pm$6\\
SPT-CL J0501$-$4455 & 75.2932 & $-44.9270$ & 4.59 & 0.25 & 0.433$\pm$0.010 & $3.80^{+0.65}_{-0.85}$ & 80$\pm$5\\
\enddata
\tablecomments{SPT-SZ clusters newly confirmed using the RM algorithm in ``scanning mode'', see Section \ref{subsec:desmatch}.  When overlapping, column entries match Table \ref{tab:sampletable}.}
\end{deluxetable*}
\end{center}

\end{document}